\documentclass[aps, prl, twocolumn, superscriptaddress, longbibliography, 10pt]{revtex4-2}

\usepackage{graphicx} 

\usepackage[caption=false]{subfig}

\usepackage{amsmath,amssymb}
\usepackage{ulem}

\usepackage[colorlinks=true, allcolors=blue]{hyperref}
\usepackage[dvipsnames]{xcolor}

\usepackage{csquotes}

\newcommand{\pheliqs}{Univ. Grenoble Alpes, CEA, Grenoble INP, IRIG, PHELIQS, F-38000 Grenoble, France}
\newcommand{\neel}{Univ. Grenoble Alpes, CNRS, Institut Néel, F-38000 Grenoble, France}
\newcommand{\lncmiG}{Univ. Grenoble Alpes, INSA Toulouse, Univ. Toulouse Paul Sabatier, EMFL, CNRS, LNCMI, F-38000 Grenoble, France}
\newcommand{\imr}{Institute for Materials Research, Tohoku University, Oarai, Ibaraki, 311-1313, Japan}
\begin{document}

\title{Bulk signatures of re-entrant superconductivity in {UTe$_2$} from ultrasound measurements}

\author{N.~Marquardt}
\email[E-mail me at: ]{nils.marquardt@lncmi.cnrs.fr}
\affiliation{\pheliqs}
\affiliation{\lncmiG}
\author{C.~Duffy}
\affiliation{\lncmiG}
\author{C.~Proust}
\affiliation{\lncmiG}
\author{S.~Badoux}
\affiliation{\lncmiG}
\author{M.~Amano Patino}
\affiliation{\pheliqs}
\affiliation{\neel}
\author{G.~Lapertot}
\affiliation{\pheliqs}
\author{D.~Aoki}
\affiliation{\imr}
\author{J.-P.~Brison}
\affiliation{\pheliqs}
\author{G.~Knebel}
\email[E-mail me at: ]{georg.knebel@cea.fr}
\affiliation{\pheliqs}
\author{D.~LeBoeuf}
\email[E-mail me at: ]{david.leboeuf@lncmi.cnrs.fr}
\affiliation{\lncmiG}

\date{\today}

\begin{abstract}
   We report bulk ultrasound measurements up to 80 T and down to 0.5 K of the field re-entrant superconducting phase of the unconventional superconductor UTe$_2$. Clear bulk signatures of superconductivity are observed in the longitudinal $c_{11}$ elastic mode for fields applied at a tilt angle of $\theta_{b-c} =30^\circ$ from $b$-axis. We confirm an upper critical field of $H_{\rm c2}\approx65$~T at 0.5~K and bulk superconductivity which survives up to $T\approx 2$~K for fields above the metamagnetic transition. The $c_{11}$ mode has propagation and displacement vectors along the $a$-axis, and for fields applied at a tilt angle of $\theta_{b-c} =30^\circ$, this mode is sensitive to the elasticity of the vortex lattice. The anomalies observed in $c_{11}$ are in part reminiscent of superconducting vortices pinned to lattice defects. Nonetheless, an excess attenuation, with respect to the normal state, is observed throughout the entire superconducting phase, suggesting unusual vortex dynamics and pinning in the field re-entrant superconducting phase of UTe$_2$.

\end{abstract}

\maketitle

The heavy-fermion material UTe$_2$ is a prominent candidate for topological spin-triplet superconductivity. Hallmarks for this include superconducting upper critical field $H_{c2}$ exceeding the Pauli limit for all crystal directions and the occurrence of multiple superconducting phases emerging under magnetic field or high pressure \cite{ran_nearly_2019,aoki_unconventional_2019-1,aoki_unconventional_2022,lewin_review_2023}. Multiple superconducting phases have also been observed in other systems such as UPt$_3$ \cite{joynt_superconducting_2002} or CeRh$_2$As$_2$ \cite{Khim2021a}. However, UTe$_2$ has the particularity that different pairing mechanisms could be at play \cite{Rosuel_2023}. At ambient pressure, when the magnetic field is applied along the $b$-axis of the orthorhombic structure, two different superconducting phases separated by a phase transition line occur \cite{Rosuel_2023,sakai_field_2023}. Th magnetic field reinforces superconductivity of the high-field phase (SC2) above 15~T, which eventually disappears abruptly at a metamagnetic transition at $H_{\rm m}=34.5$~T \cite{knebel_field-reentrant_2019,ran_extreme_2019, miyake_metamagnetic_2019,knafo_magnetic-field-induced_2019}. While for the low-field superconducting phase (SC1) evidence for a spin-triplet pairing is found \cite{Matsumura_Large_2023a,Matsumura_b-axis_2025,li_observation_2025,Carlton_revealing_2025,Hayes_robust_2025,suetsugu_fully_2024, Gu_pair_2025}, the pairing symmetry of the high field SC2 phase is still controversial \cite{kinjo_change_2023,helm_field-induced_2024,zhang_electronic_nodate,lewin_high-field_2025}.

In the field polarized (FP) phase for $H>H_{\rm m}$, no sign of superconductivity is found $H\parallel b$. However, by tilting the magnetic field away from the $b$-axis a re-entrance of superconductivity occurs \cite{ran_extreme_2019}. This field-polarized superconducting phase (SC3) appears to be confined inside the field polarized region for $H>H_{\rm m}$ \cite{lewin_high-field_2025}, though recent studies claim it also exists in close proximity to $H_{\rm m}$ inside the paramagnetic regime \cite{wu_quantum_2025}. The underlying mechanism for this SC3 phase is not clear \cite{lewin_review_2023}. It is proposed that the off $b$-axis magnetic field leads to metamagnetic quantum criticality which may stabilize SC3 in UTe$_2$ \cite{wu_quantum_2025}. Alternatively, the compensation of the applied field by an exchange field (Jaccarino-Peter effect) is also discussed \cite{helm_field-induced_2024}.

The SC3 phase remains mostly studied with transport techniques, namely contactless surface conductance measurements \cite{lewin_high-field_2025,wu_quantum_2025,schonemann_sudden_2023,wu_superconducting_2025}, resistivity \cite{ran_extreme_2019,knafo_comparison_2021,thebault_field-induced_2025,helm_field-induced_2024,aoki_high_2024,Frank2024,Lewin2024} and Hall resistivity \cite{helm_field-induced_2024}. Up to now, bulk detection of the SC3 phase are scarce. 
The magnetocaloric effect has been measured in the SC3 phase \cite{schonemann_sudden_2023}. It shows reversibility in the SC3 regime which is interpreted as a consequence of the opening of a bulk superconducting gap \cite{schonemann_sudden_2023}. 

\begin{figure}[tb]
 	\includegraphics[width=\linewidth]{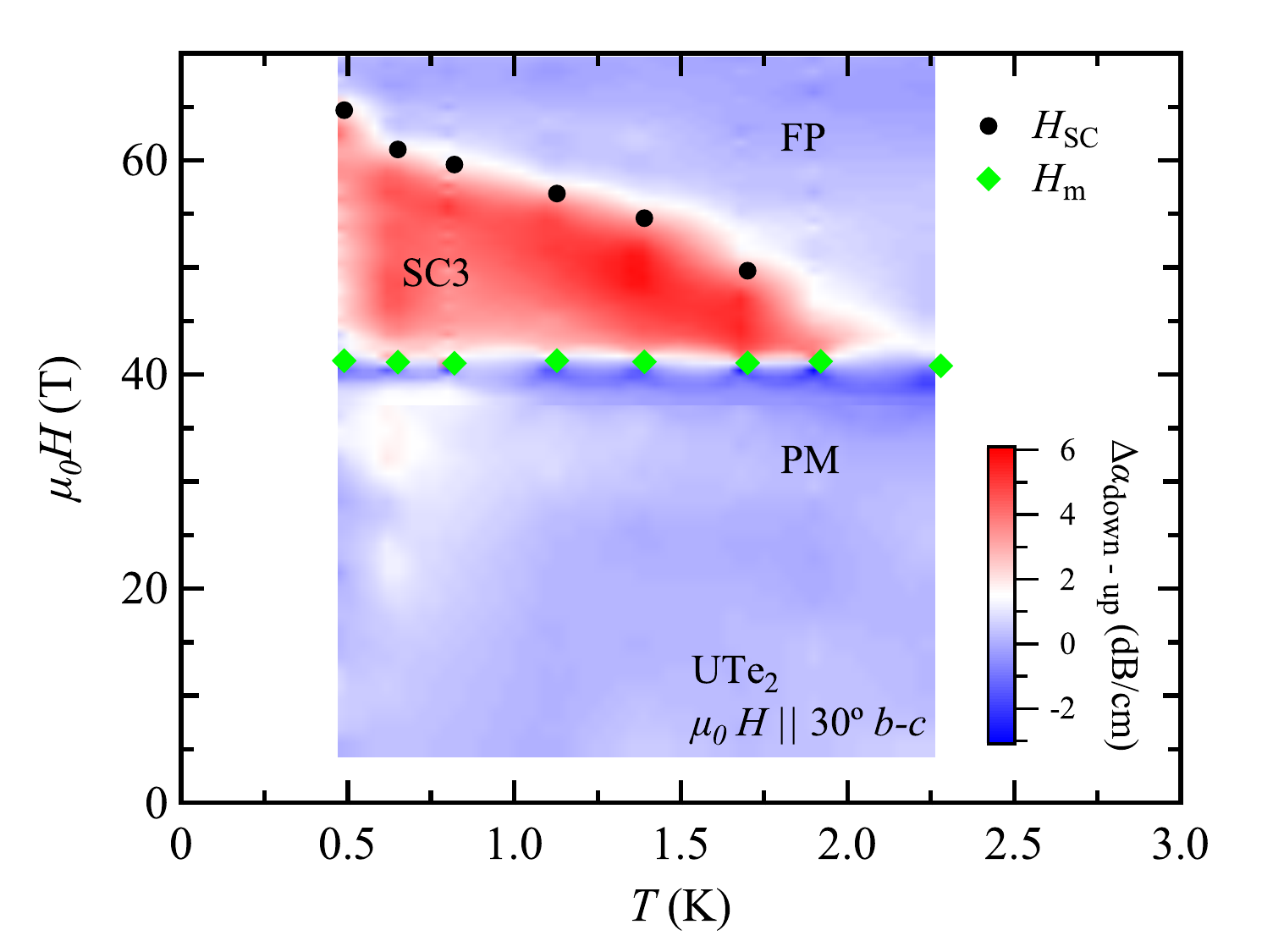}
 	\caption{Phase diagram of SC3 phase of UTe\textsubscript{2} for $H$ applied with $\theta_{b-c} =30^\circ$. The metamagnetic transition $H_{\rm m}$ is shown in green diamonds and $H_{\rm SC}$ is shown in black circles (see sound velocity data in Fig. \ref{c11_SC}a). The color code represents the amplitude of the hysteresis loop in the sound attenuation $\Delta \alpha$ shown in Fig. S13 in Ref.~[28].}
 	\label{Phase}
 \end{figure}

In this Letter we present an ultrasound study of the field re-entrant superconducting phase of UTe$_2$. We measured the SC3 phase of UTe$_2$ in pulsed fields up to 80 T and down to 0.5 K. We observe anomalies in the sound velocity and in the attenuation of the longitudinal elastic mode $c_{11}$. We observe an hysteresis, showing excess attenuation and lattice stiffness (w.r.t. the normal state) on the down-sweep part of the magnetic field pulse. In the field descending curve this hysteresis opens at a $T$-dependent critical field  $H_{\rm SC}$, that correlates with $H_{\rm c2}$ measured with other probes, and extends to the metamagnetic transition $H_{\rm m}$. Figure~\ref{Phase} displays the bulk $H-T$ phase diagram determined from our ultrasound measurements, which highlights the SC3 phase. 
We attribute the ultrasound anomalies to the presence of superconducting vortices pinned to lattice defects. This implies the existence of a bulk (mixed) superconducting state. Remarkably, a large residual ultrasound attenuation is found across the entire superconducting phase. This behaviour suggests non-trivial vortex pinning and dynamics.

 \begin{figure}[b]
 \includegraphics[width=\linewidth]{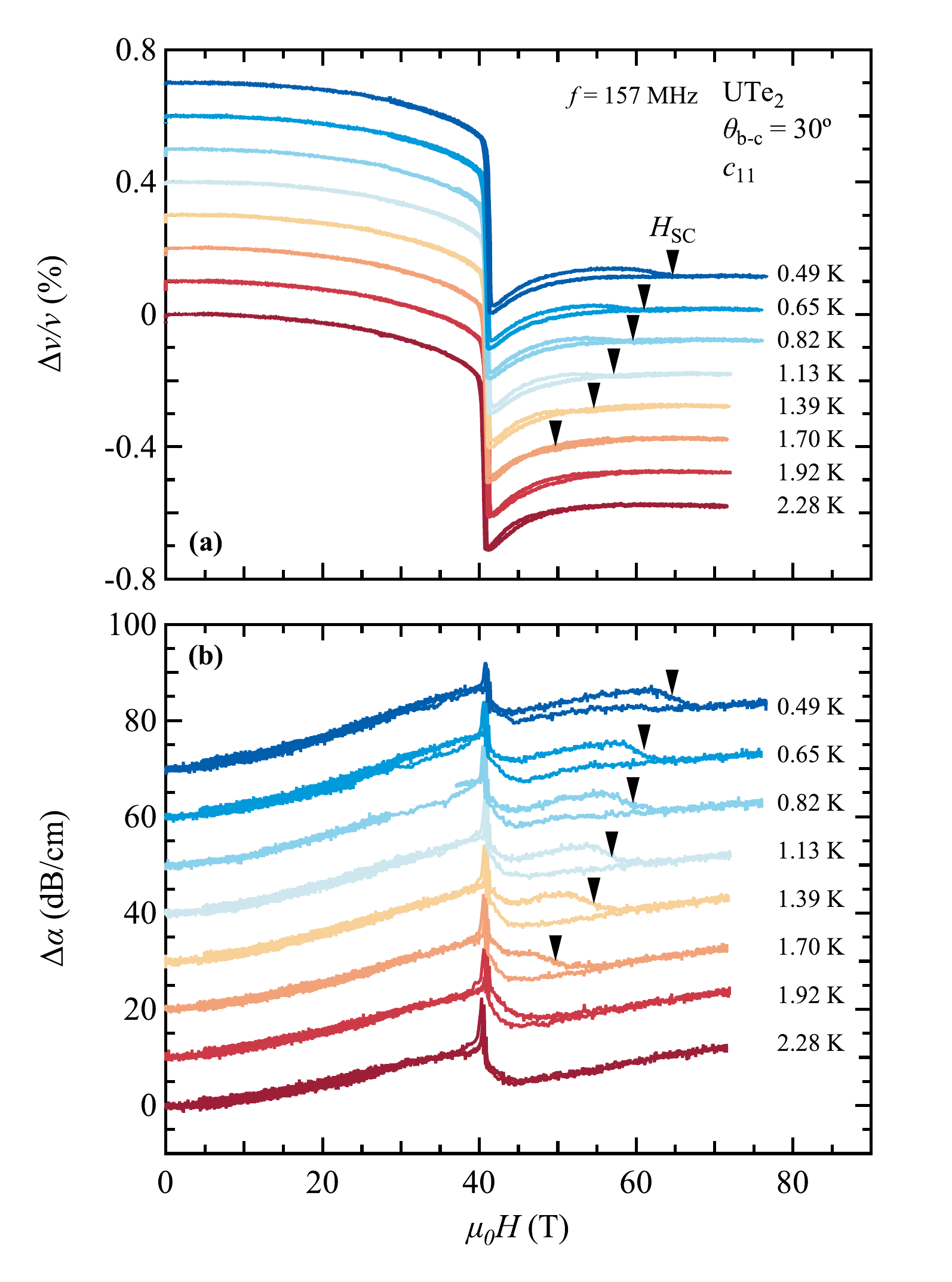}
 	\caption{(a) Field dependent relative change of sound velocity ${\Delta v}/{v}$ in the $c_{11}$ mode for a field orientation of $\theta_{\rm b-c}=30^\circ$ for temperatures ranging from $0.49-2.28$~K. Curves are shifted vertically for clarity. $H_{\rm SC}$ (fig. \ref{Phase}) is a kink in the field decreasing curve shown by black arrows. Kinks is best observed in Fig. S12 in Ref. [28]. The corresponding field dependent attenuation change $\Delta \alpha$ is displayed in panel (b) with the same color code and shifted vertically for clarity. The black arrow indicates the same magnetic field value as in (a). The ultrasound frequency is $f= 157$~MHz. Changes in velocity and attenuation are calculated with respect to the zero field value.}
	\label{c11_SC}
\end{figure}

We use the pulse-echo ultrasound method to measure changes in sound velocity and attenuation in different elastic modes in single crystals of UTe$_2$ with a $T_{\rm c} \approx 2$~K (see Supplemental Material \cite{SuppMat} for methods and sample characterization). Figure~\ref{c11_SC} shows ultrasound data in the $c_{11}$ mode of UTe$_2$ for fields applied at a tilt angle of $\theta_{\rm b-c}=30^\circ$ from the $b-$axis in the $(b,c)$ plane. The $c_{11}$ mode corresponds to an acoustic wave with propagation $k\parallel a$ and polarization $u\parallel a$. The sound velocity $\Delta v/v$ (Fig.~\ref{c11_SC}a) decreases with increasing field up to $\mu_0H_{\rm m}  \approx 41$~T. At $\mu_0H_{\rm m}$ a large step-like  anomaly signaling the metamagnetic transition occurs, confirming our tilt angle $\theta_{\rm b-c}$. The overall behaviour for $H\leq H_{\rm m}$ is similar to that for a field applied along the $b-$axis ($\theta_{\rm b-c}=0^\circ$), as seen in Fig.~\ref{c11 and c66}a (see below), and as shown previously in other acoustic modes \cite{Valiska_2024}. The simultaneously measured sound attenuation $\Delta \alpha$ (see Fig.~\ref{c11_SC}b) increases with increasing field for $H<H_{\rm m}$, and shows a sharp peak at $H_{\rm m}$. Across the temperature range studied here, the ultrasound anomalies associated with the metamagnetic transition are essentially temperature independent. Above $H_{\rm m}$, and for $T=2.28$~K, the ultrasound properties are featureless. Note that the up and down sweep of the magnetic field pulse are both shown in Fig.~\ref{c11_SC}. For $T = 2.28$~K, both sweeps essentially superimpose across the whole field range, indicating overall good isothermal conditions, except at or near $H_{\rm m}$. Indeed, a small difference in $\Delta v/v $ is measured inside the normal state FP region, and will be discussed later in the manuscript. 

\begin{figure*}
    \includegraphics[width=1\linewidth]{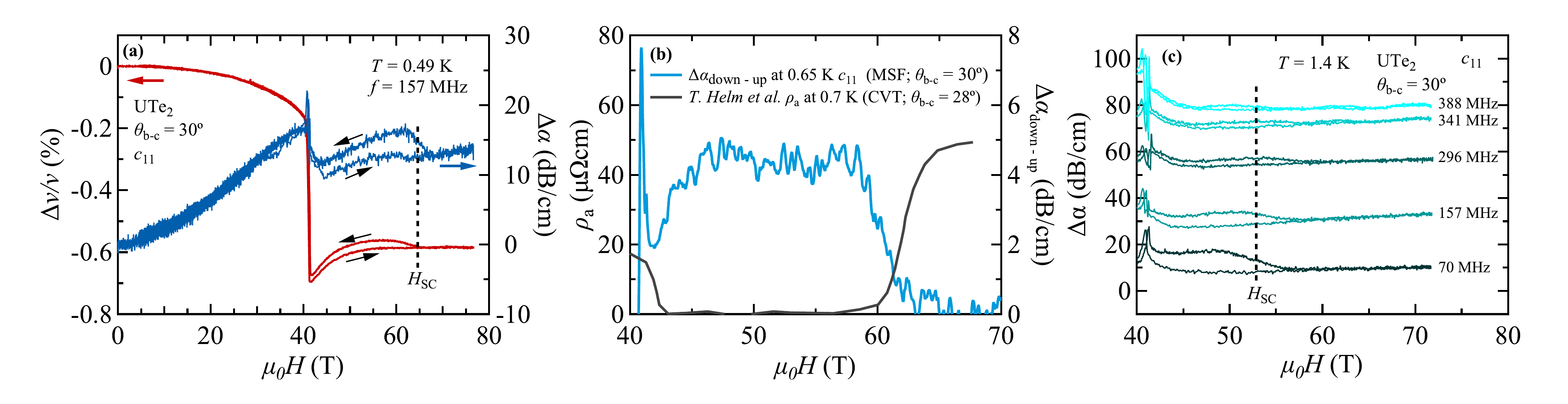}
 	\caption{a) Field dependence of $\Delta v/v$ (red, left axis) and $\Delta \alpha$ (blue, right axis), for ascending and descending field, measured in the $c_{11}$ mode at $T=0.49$~K. The dashed line indicates $H_{\rm SC}$ and black arrows indicate increasing or decreasing field. b) Hysteresis of sound attenuation (blue) as a function of field for $T=0.65$~K in the $c_{11}$ mode. The blue curve is obtained by calculating ($\Delta \alpha_{\rm down} - \Delta \alpha_{\rm up})$, see in Fig. S13 in Ref.~[28]. The grey line shows the resistivity, $\rho_a$, measured at similar angle and temperature, from Ref. \cite{helm_field-induced_2024}. c) $\Delta \alpha$ for frequencies ranging from 70 to 388~MHz at a temperature $T\approx1.4$~K, shifted vertically for clarity. $H_{\rm SC}$ measured at 157~MHz is indicated with the dashed line.}
 	\label{pannel}
 \end{figure*}

For $T\leq 1.7$~K a new anomaly appears in the ultrasound properties of the $c_{11}$ mode. An hysteresis loop opens between the field-up and down sweeps, both in the sound velocity and attenuation. The hysteresis is best seen in Fig.~\ref{pannel}a, where the up- and down-sweeps are shown for $T=0.49$~K. The down-sweep shows the largest anomalies, with an increase in $\Delta \alpha$ and in $\Delta v/v$ w.r.t the high-field normal state. Anomalies are also observed, though with smaller amplitudes, on the up-sweep (see Fig.~\ref{pannel}a and Fig.~S11 and S12 \cite{SuppMat}). The kink in the down-sweep of $\Delta v/v$, labelled $H_{\rm SC}$ in Fig.~\ref{pannel}a, corresponds to an inflexion point in $\Delta \alpha$. $H_{\rm SC}$ is shown with black arrows in Fig.~\ref{c11_SC}a,b. The amplitude of the hysteresis is found to be frequency dependent, as shown in Fig. \ref{pannel}c and Fig. S2 \cite{SuppMat}. The anomalies appear at higher fields as temperature is decreased, and it can be tracked from $T\sim 1.70$~K down to the lowest measurement temperature $T=0.49$~K. It can be analysed by plotting the difference between the up- and down-sweep part of the curves, as done in Fig.~\ref{pannel}b for $T=0.65$~K and in Fig.~S13 for all temperatures studied here. The hysteresis difference in the attenuation is plotted in the false colour plot phase diagram of Fig.~\ref{Phase}, along with the high field anomaly $H_{\rm SC}$ in $\Delta v/v$ (black arrows in Fig.~\ref{c11_SC}a) and $H_{\rm m}$. Blue regions indicate no difference between up and down-sweeps, while red region indicates hysteresis with $\Delta \alpha_{\rm down} - \Delta \alpha_{\rm up}> 0$. 
The red region of the phase diagram of Fig.~\ref{Phase} tracks remarkably well the phase boundary of the field re-entrant superconducting phase of UTe$_2$ found near this tilt angle (see \cite{wu_superconducting_2025,wu_quantum_2025} and Fig. S9 \cite{SuppMat}). This correlation indicates that the anomalies found in the $c_{11}$ mode are a bulk signature of the field re-entrant superconducting phase of UTe$_2$. This is the main result of this study. The existence of bulk signatures of the SC3 phase indicates that superconductivity itself is a bulk phenomenon, thus excluding local, extrinsic and minority effects.

Before dwelling into the possible origins of the excess attenuation and elasticity found in the SC3 phase, a discussion of possible sample heating effects is required. In particular, determine to which extent a temperature cycle could be responsible for the hysteresis found in the ultrasound properties inside the superconducting phase.
One possible source of heating comes from the rapidly varying field found in pulsed magnets. Nonetheless, our analysis shown in Supplemental Material \cite{SuppMat} indicates the hysteresis found inside the SC3 phase does not originate from eddy current heating of the sample (see Fig.~S6 \cite{SuppMat}).
Another source of sample heating is the metamagnetic transition \cite{schonemann_sudden_2023}. For $T>T_{\rm c}$, the magnetocaloric effect at $H_{\rm m}$ results in a temperature increase of the sample \cite{miyake_2021}, producing an irreversible sample temperature cycle in the FP region \cite{schonemann_sudden_2023}. This could explain the small hysteresis observed at $T= 2.28$~K in the sound velocity. Note that the hysteresis at $T=2.28$~K is inverted with that observed for $T\leq T_{\rm c}$ (see Fig.~\ref{c11_SC} and S10 \cite{SuppMat}) and absent in the attenuation. In the Supplemental Material \cite{SuppMat}, we discuss in detail this magnetocaloric effect. While we cannot fully rule out that part of the measured hysteresis is due to a temperature cycling, it is unlikely to cause a hysteresis, which amplitude is frequency dependent, and which opens at a $T-$dependent onset field that closely tracks $H_{\rm c2}(T)$ as determined with other techniques in different conditions (see Fig. S9 \cite{SuppMat}). Only in the extreme case where the sample heats up above $T_{\rm c}$ when crossing $H_{\rm m}$ on the upsweep, and somehow rapidly cools down to bath temperature before the downsweep begins, could we account for a smaller anomaly found in the upsweep than on the downsweep. Nevertheless, an independent issue remains: the origin of the anomaly itself. SC3 features an excess attenuation and elastic stiffness on the downsweep. These features that cannot be explained by temperature cycling, are adressed in the following discussion.

Acoustic anomalies arising from superconductivity can have several origins. First, a coupling between the superconducting order parameter and strain can produce elastic anomalies. In UTe$_2$ these anomalies have been well studied in zero magnetic field \cite{theuss_resonant_2024,theuss_2024}. This coupling can lead to an increase in the sound velocity below $T\leq T_{\rm c}$ and step-like anomaly at $T=T\leq T_{\rm c}$ in longitudinal modes such as $c_{11}$. However, the sound velocity anomaly measured in UTe$_2$ at zero field in $c_{11}$ is an order of magnitude smaller than what is measured here \cite{theuss_2024}, suggesting a different mechanism for the anomalies found here in the SC3 phase. Moreover, the opening of the superconducting gap usually causes a decrease of the sound attenuation inside the superconducting state \cite{Bardeen_Theory_1957}. Here we observe the opposite. Compared to the normal state, an excess sound attenuation occurs throughout the entire superconducting phase. This phenomena, which is non-trivial as we will see, should arise from an additional dynamical process which occurs specifically inside the SC3 phase. One possible origin of attenuation is Landau-Khalatnikov (LK) relaxation of the superconducting order parameter below $T_{\rm c}$ (see \cite{miyake_landau-khalatnikov_1986} and \cite{SuppMat} for more details). While the model shows a fair agreement with the attenuation data (see Fig.~S4 \cite{SuppMat}), several key aspects disqualifies it. It does not reproduce the observed frequency dependence and it does not produce hysteresis by itself, unless the relaxation time is longer than the duration of the pulse. However, fit parameters indicate a relaxation time $\tau \sim 10^{-10}$~s (see Fig.~S4 \cite{SuppMat}), much smaller than the duration of the pulse: $\Delta t \sim 16$ ms for the whole FP regime. Finally, such relaxation process usually produces a decrease of the sound velocity for $T=T\leq T_{\rm c}$, while an increase is observed here.

The second possible explanation for the ultrasound anomalies found inside the SC3 phase of UTe$_2$ rely on a coupling between superconducting vortices and the crystal lattice, introducing another type of dynamical process. This coupling occurs if vortex pinning centers exist in the system. Vortex pinning produces irreversible behaviour in the magnetostriction \cite{eremenko_magnetostriction_1999} or ultrasound properties \cite{Campbell_2022}, when the field is applied below superconducting $T_{\rm c}$. Consequently, this scenario could naturally explain the hysteresis observed here. 
Because the pulse-echo ultrasound technique is also sensitive to changes in the length of the sample, the anomaly measured here could be due to irreversible magnetostriction $\Delta L_a$, with $L_a$ the sample's length along the propagation direction $k\parallel a$. This is proposed to be the origin of a hysteresis in $\Delta v/v$ in CeRu$_2$ \cite{Wolf_1997}. Irreversibility due to $\Delta L_a (\mu_0 H)$ should be observed for all modes propagating along the $a-$axis, such as $c_{66}$ which is a transverse mode with propagation direction $k\parallel a$ and polarisation $u\parallel b$. The field dependence of $c_{66}$ is shown in Fig.~\ref{c11 and c66}b. It shows a softening prior to $H_{\rm m}$ and large positive step ($\Delta v/v \sim + 10^{-2}$) at $H_{\rm m}$, as observed previously for $\theta_{\rm b-c}=0^\circ$ \cite{Valiska_2024}. For $\theta_{\rm b-c}=30^\circ$, we do not observe any hysteresis in $c_{66}$ for $H > H_{\rm m}$ and $T<T_{\rm c}$, in contrast to $c_{11}$ at similar angle and comparable frequency. As both modes would be sensitive in comparable magnitude to the length change, this rules out the magnetostriction as a cause for the anomaly observed in $c_{11}$.

 \begin{figure}[t]
 \includegraphics[width=\linewidth]{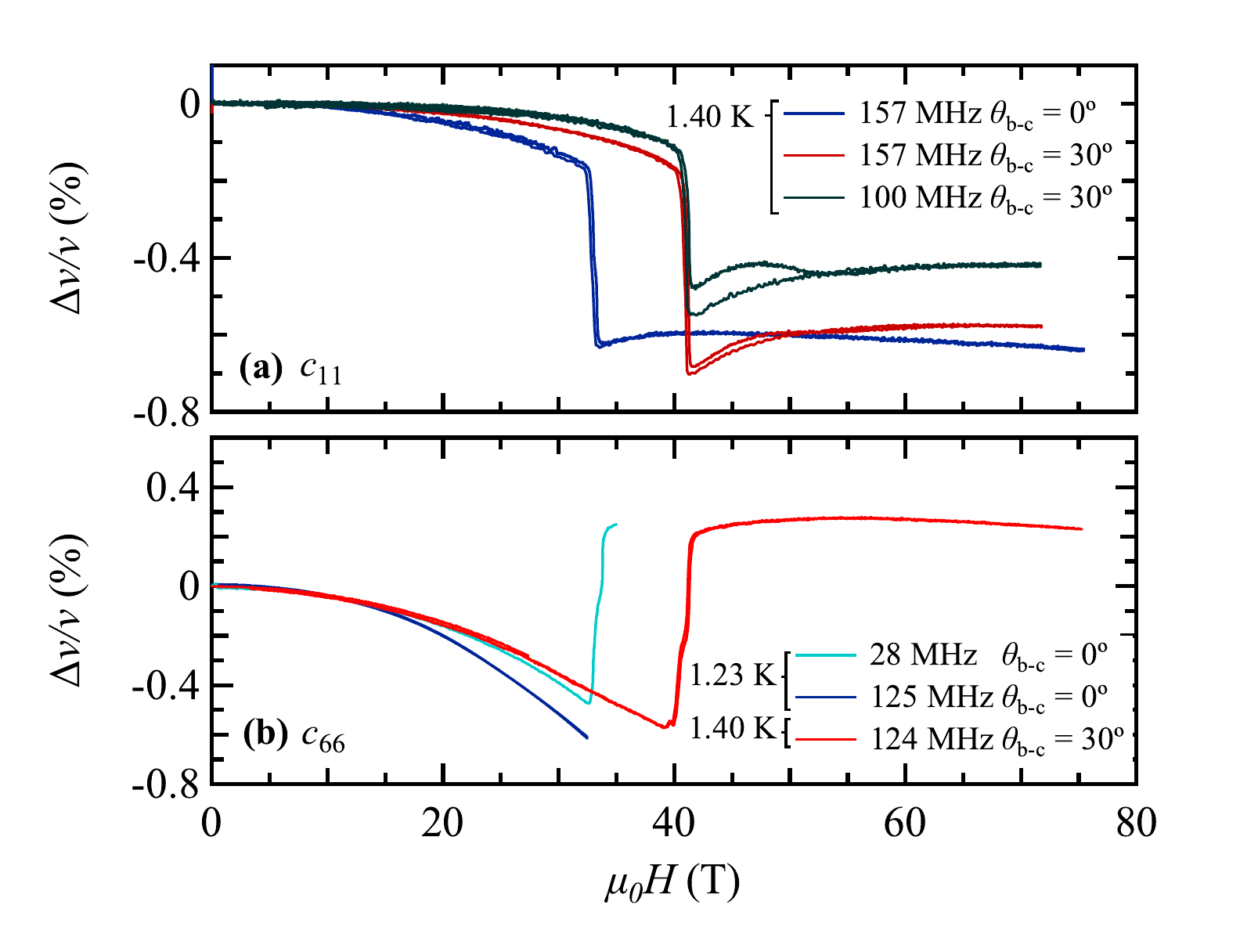}
 	\caption{The $c_{11}$ and $c_{66}$ mode relative sound velocity are compared at various ultrasound frequencies and for field orientations along the $b$-axis and for  $\theta_{\rm b-c}=30^\circ$ for $T=1.4$~K.}
 	\label{c11 and c66}
\end{figure}

The ultrasounds can also be sensitive to the elastic stiffness and dynamics of the vortex lattice. When the vortex lattice is pinned, the sound velocity $v$ has an additional contribution $\Delta c_{ij}^v$ from the vortex lattice : $\rho v^2 =c_{ij}^c+\Delta c_{ij}^v$, with $\rho$ the mass density, and $c_{ij}^c$ the crystal lattice elastic modulus. A triangular vortex lattice has three independent elastic moduli: a compression modulus $c_L^v=c_{11}^v-c_{66}^v$, a tilt modulus $c_{44}^v$ and a shear modulus $c_{66}^v$ (see \cite{Brandt_1995,Campbell_1972} for reviews). For $k\parallel u\perp H$ ($c_{11}$ mode, Figs. \ref{c11_SC} and \ref{c11 and c66}a), we probe the compression modulus $c_L^v$ of the vortex lattice. For the $c_{66}$ mode, some component of the displacement vector $u$ is perpendicular to the field. Hence, the $c_{66}$ mode should be sensitive to $c_{66}^v$. Since, in high fields, and for a Ginzburg-Landau parameter $\kappa \gg 1$ \cite{Rosuel_2023}, we have $c_{66}^v\ll c_{11}^v=c_{44}^v$ \cite{Campbell_1972}, this scenario relying on the dynamics of pinned vortices naturally explains the absence of any resolved anomalies in $c_{66}$ (see Fig.~\ref{c11 and c66}b). It also accounts for the fact that the elastic constant is larger on the down sweep than on the up sweep (see Figs.~\ref{c11_SC}, \ref{pannel} and S13). Indeed, assuming a critical state, and recalling that $c_{11} \approx B^2/4\pi$ \cite{Campbell_1972}, it is expected that the pinned vortex lattice will have higher elastic stiffness when entering the superconducting state from the FP state at $H_{\rm SC}$ (decreasing field) than at $H_{\rm m}$ (increasing field). A phenomenological model based on thermally assisted flux flow (TAFF) \cite{Kes_1989} can be used to reproduce the temperature dependence of the vortex lattice contribution to the sound velocity $\Delta c_{ij}^v$. Within this model, $\Delta c_{ij}^v$ is given by the vortex lattice elastic modulus $c_{ij}^v$, renormalized by the vortex lattice dynamics associated with the TAFF \cite{Pankert_1990,Wolf_1996,Wolf_1997}. The model captures qualitatively the temperature dependence of the sound velocity below $T_{\rm c}$, as shown in Fig.~S5 \cite{SuppMat}. 

All in all, the sound velocity data suggest that vortices contribute to the ultrasound properties inside the SC3 phase. Nonetheless, some part of the data remain difficult to capture within this approach. First the vortex elastic constant expected at 50~T is $c_{11}^v\approx 2$~GPa. With a lattice elastic modulus $c_{11}^c=81$~GPa \cite{theuss_resonant_2024}, it implies an expected increase $\Delta v/v\sim 10^{-2}$, almost 30 times larger than the value $\Delta v/v\sim 4\times10^{-4}$ measured at low $T$ at 157~MHz (see Fig.~S5). Moreover, the sound velocity anomaly, approximated to the size of hysteresis, strongly depends on the measurement frequency. As can be seen in Fig.~\ref{pannel}c and Fig.~S2, the size of the anomaly decreases rapidly with increasing frequency for both the attenuation and sound velocity. In contrast the TAFF model actually predicts a sound velocity anomaly which is independent of frequency, and the anomaly in the attenuation should become larger with increasing frequency. More surprises arise when scrutinizing the ultrasound attenuation. Vortices will produce attenuation only when they are allowed to move, meaning close to the irreversibility field or $H_{\rm c2}$. Here we find excess attenuation (red area in Fig.~\ref{Phase}) throughout the entire superconducting phase. This is best illustrated by comparing our attenuation data with resistivity $\rho$ data at similar tilt angle and temperature, as done in Fig.~\ref{pannel}b. Throughout the entire $\rho \approx 0$ state, the attenuation is larger than in the normal state. 

In order to explain this behaviour within a TAFF model, a large distribution of pinning energies has to be invoked, implying inhomogeneous pinning barriers. 
Alternatively, viscous motions of vortices could occur close to $H_{\rm c2}$ and to $H_{\rm m}$, hence producing attenuation throughout the entire superconducting phase when these two regimes overlap. In fact, the attenuation data suggest the existence of two overlapping peaks (see Fig.~\ref{pannel}b and S13). Still, the frequency dependence of both the sound velocity and attenuation suggests that low frequency sound waves couple more efficiently with vortices. %
An inhomogeneous superconducting phase, where the strain couples to a vortex lattice locked-in with the spatially modulated superconducting order parameter \cite{Watanabe_2004,Imajo_2022}, could potentially explain some of the features found here. Anisotropic pinning barriers are suggested to occur within the field re-enforced superconducting state SC2 \cite{zhang_electronic_nodate}. A similar type of modulation could occur if the superconductivity at such high fields is due to the Jaccarino Peter effect \cite{helm_field-induced_2024}. If the SC3 phase is spin-singlet, a FFLO-state could exist. The associated modulated superconducting order parameter would introduce pinning planes with a changing $q$-vector in magnetic field and therefore result in a change of the pinning.

While the detailed origin of the ultrasound signatures of the SC3 remain unclear at this stage, a signal arising from vortex dynamics seems most likely. Measurements of the critical current inside the SC3 phase of UTe$_2$ would be helpful in order to improve our understanding of the vortex dynamics. In any case, the behaviour found here contrasts with what we observe in the other two, ambient pressure, superconducting phases of UTe$_2$ found below $H_{\rm m}$. Within our resolution we do not observe any superconducting contribution of the lower field superconducting phases inside the paramagnetic region, suggesting a smaller coupling of the $c_{11}$ mode to the vortices found in SC1 and SC2 phases. This contrast suggests that vortex pinning is different in the field re-entrant SC3 phase. 

In summary, we have conducted an ultrasound study of UTe$_2$, measuring different elastic modes at different tilt angles in pulsed magnetic fields. For $\theta_{\rm b-c}=30^\circ$, we observe a bulk signature of the SC3 phase in the sound velocity and attenuation. The anomalies are best explained by pinning of superconducting vortices. . 
Nonetheless, several features remain puzzling. An excess attenuation is observed throughout the entire SC3 phase, and a strong frequency dependence is observed. Both observations suggest unconventional vortex dynamics in the field re-entrant superconducting phase of UTe$_2$ and should be addressed by future experiments.

\begin{acknowledgments}
We are grateful for many stimulating discussions with D. Braithwaite, A. Pourret, T. Vasina, M. Zhitomirsky, J. Flouquet, Y. Yanase, A. Miyake, T. Klein, H. Suderow and B. Ramshaw. We acknowledge experimental support by D. Vignolles and N. Bruyant. 
We acknowledge financial support from the French National Agency for Research ANR within the project FRESCO No. ANR-20-CE30-0020, SCATE No. ANR-22-CE30-0040 and from the JSPS programs KAKENHI (JP19H00646, JP20K20889, JP20H00130, JP20KK0061, JP20K03854, JP22H04933). We acknowledge support of the LNCMI-CNRS, member of the European Magnetic Field Laboratory (EMFL), and from the Laboratoire d’excellence LANEF (ANR-10-LABX-0051). 
\end{acknowledgments}

\bibliography{biblio.bib}

\clearpage
\setcounter{figure}{0}
\renewcommand{\thefigure}{S\arabic{figure}}
\section{Supplementary Materials for "Bulk signatures of re-entrant superconductivity in UTe$_2$ from ultrasound measurements"}

\section{Methods and Samples}
\begin{figure}[h]
 	\includegraphics[width=0.98\linewidth]{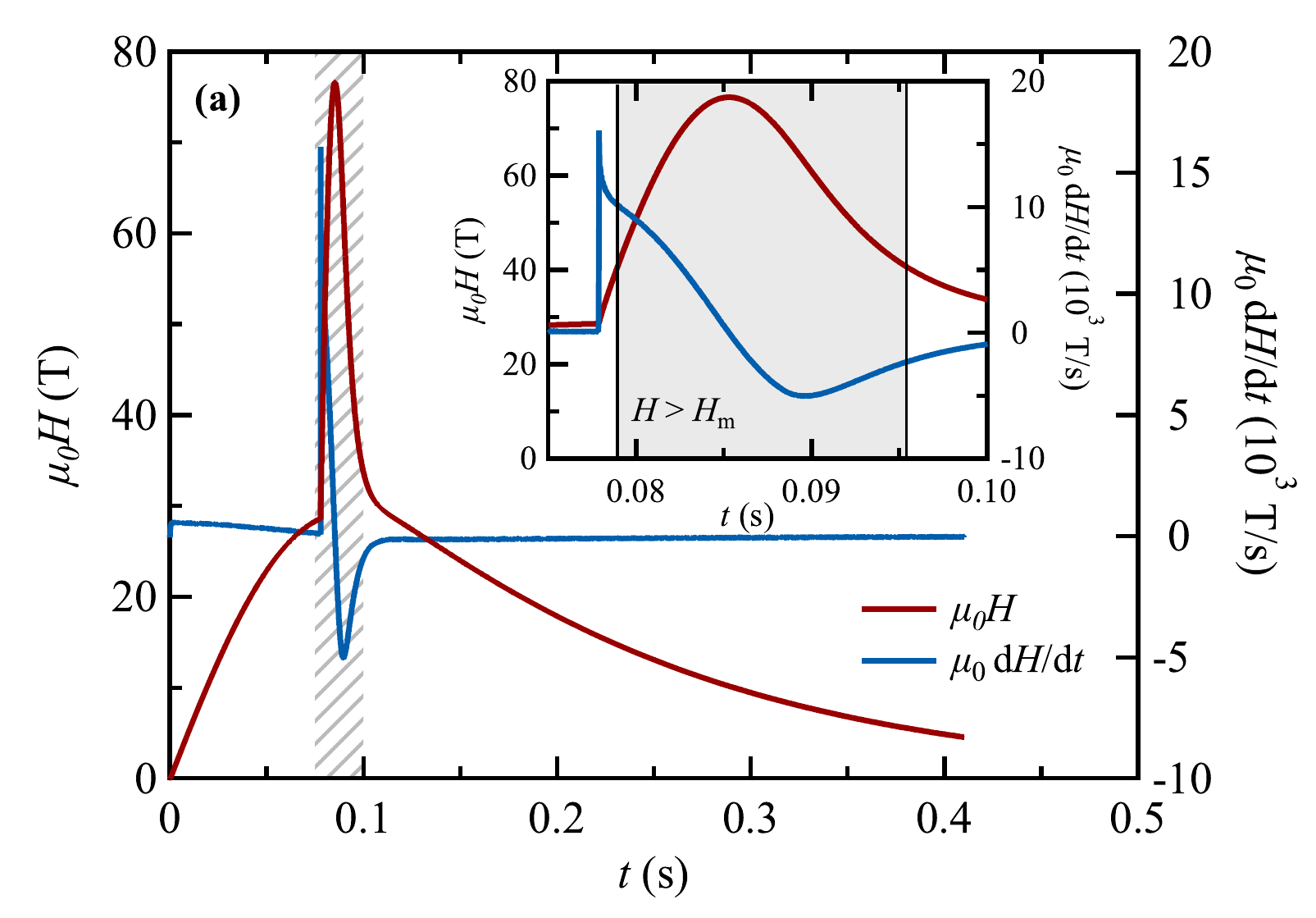}
 	\hfill
    \includegraphics[width=1\linewidth]{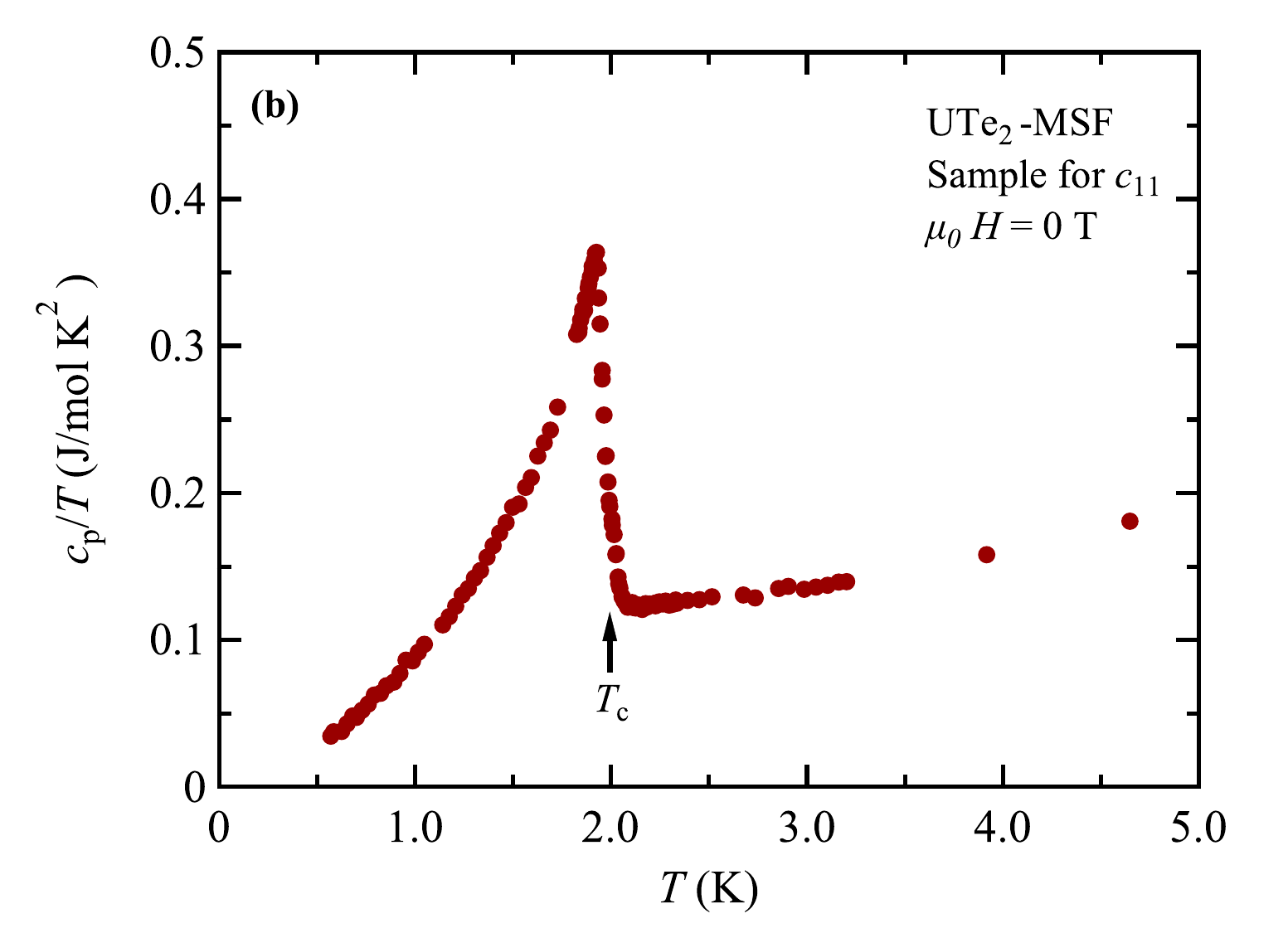}
	\caption{a) Magnetic filed profile of the dual-coil magnet used for pulsed field measurements up to 80 T. Red curve shows magnetic field versus time, and blue curve shows time-derivative of magnetic field $dB/dt$. Inset shows a zoom on the pulse of the inner coil. b) Zero field specific heat measured through superconducting $T_{\rm c}\approx 2$ K (shown by an arrow) and plotted as $c_p/T(T)$ for the sample where the $c_{11}$ mode was measured.}
	\label{specific_heat_field}
\end{figure}

The pulse echo ultrasound method is used to measure the relative change of sound velocity ${\Delta v}/{v}$ and attenuation ${\Delta \alpha}$ simultaneously. For high symmetry modes, the relative change of the elastic constants $c_{ij}$, here in Voigt notation, is by a factor of two proportional to the relative sound velocity change ${\Delta c_{ij}}/{c_{ij}}=2{\Delta v}/{v}$. We use LiNbO$_3$ transducers that are glued on polished and oriented single crystals. We measure in two different system either working in liquid \textsuperscript{4}He or \textsuperscript{3}He environment to achieve stable thermal conditions down to 450~mK. Measurements up to 36 T are carried out in DC resistive magnets of the LNCMI site in Grenoble. Higher fields up to 80 T were obtained at the LNCMI's pulsed field facility in Toulouse, using a dual-coil magnet. Its field profile is shown in Fig. \ref{specific_heat_field}. 

We measured the longitudinal $c_{11}$ and transverse $c_{66}$ crystal elastic modes. For the $c_{11}$-mode the propagation vector of the ultrasonic deformation wave is along $\Vec{k}||a$-axis and pointing in the same direction as the polarisation $\Vec{u}||a$-axis. For the transverse $c_{66}$-mode we chose the same propagation vector $\Vec{k}||a$-axis, but the polarisation is now perpendicular $\Vec{u}||b$-axis. 

We study two high quality Molten Salt Flux (MSF) samples from the same batch with a ($T_c\approx2$~K) \cite{sakai_single_2022, aoki_molten_2024}. Both samples have similar dimensions, wherefore the sample for $c_{\rm 11}$ has a length of $l_a = 1425~\mu$m along the $a$-axis and mass of $m=1.42$~mg. The samples are grown and prepared at the CEA-Grenoble. The zero field superconducting transition is verified via specific heat measurements, showing a sharp single jump at $T_{\rm c}$, as shown in Fig. \ref{specific_heat_field}. The samples are oriented via Laue diffraction and the respective field angle with respect to the crystallographic axis is achieved via a 3D-printed wedge. 

\clearpage

\section{Frequency dependence of ultrasound properties}
\begin{figure}[h]
 	\includegraphics[width=1\linewidth]{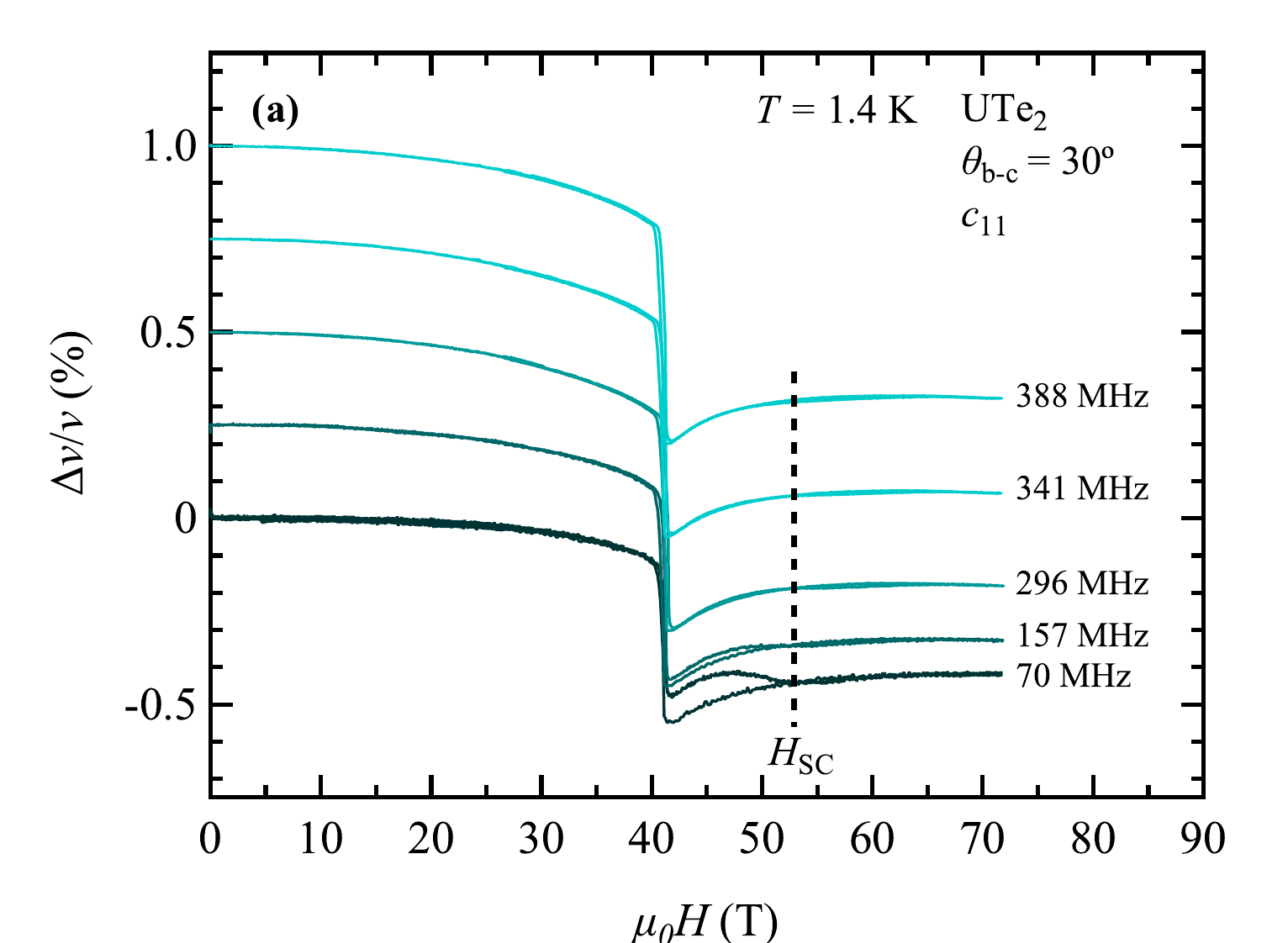}
\hfill
 	\includegraphics[width=1\linewidth]{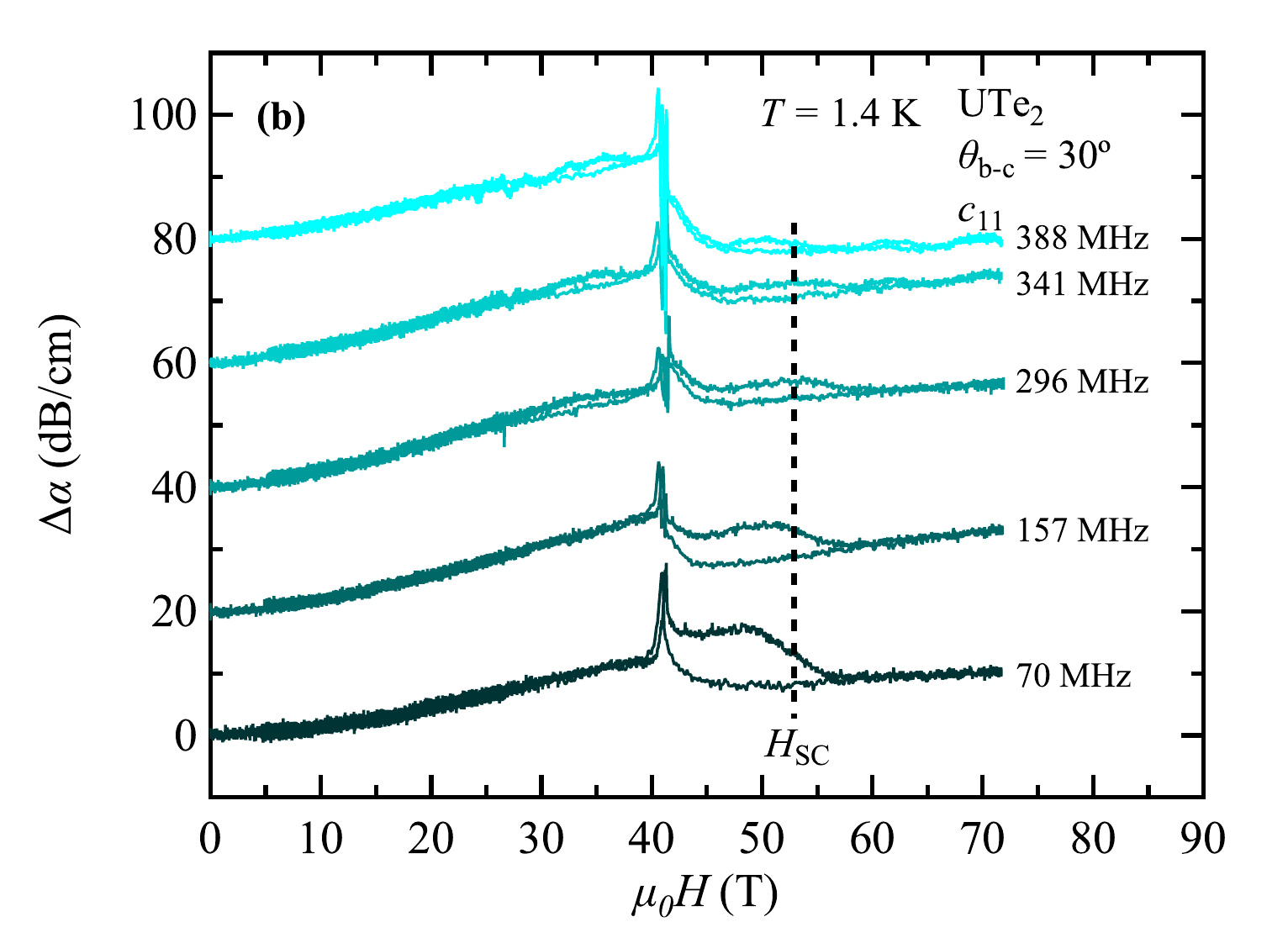}
 	\caption{A comparison of the effect of different ultrasonic frequencies ranging from 70 to 388~MHz at a temperature $T=1.4$~K on the ultrasound properties. a) Field dependence of the relative change of sound velocity with an offset of 0.25\% between each frequency. b) Field dependence of the ultrasound attenuation with an offset of 20~dB/cm.}
	\label{Stack_freq}
\end{figure}

\begin{figure}[h]
 	\includegraphics[width=1\linewidth]{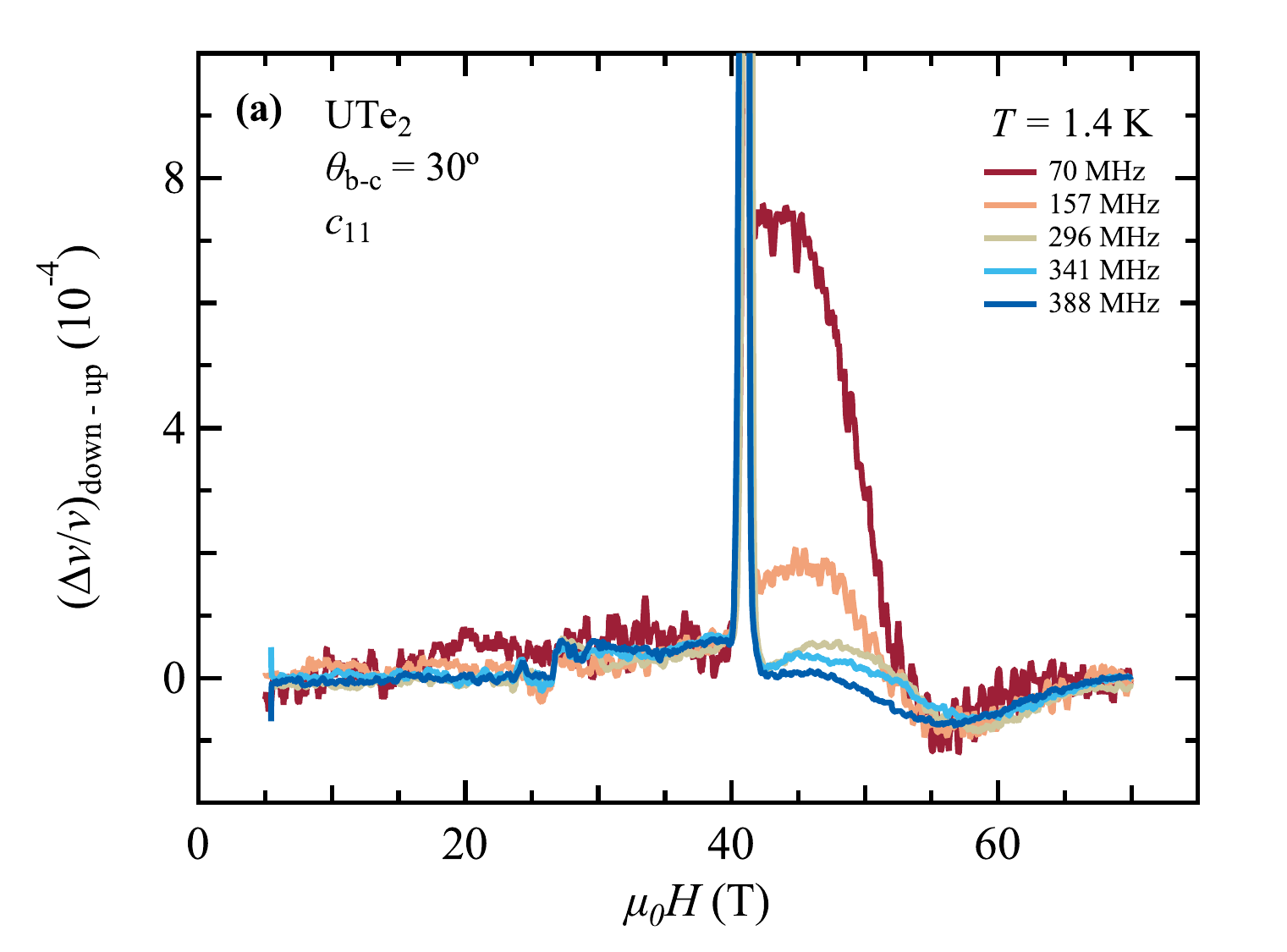}
\hfill
 	\includegraphics[width=1\linewidth]{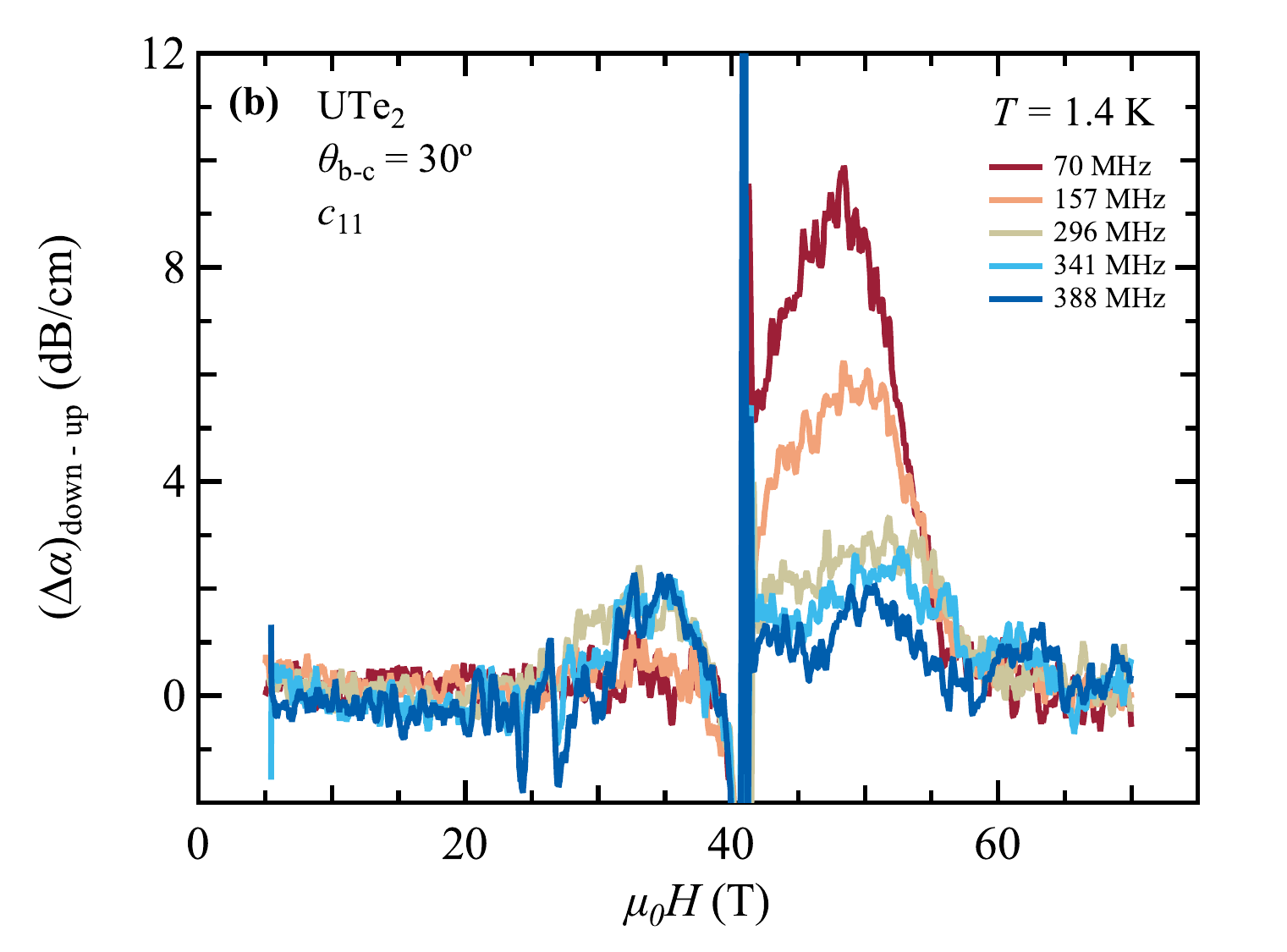}
 	\caption{Frequency dependence of the sound velocity (left panel) and attenuation (right panel) magnetic field hysteresis loop, plotted as $(\Delta v/v)_{\rm down}-(\Delta v/v)_{\rm up}$ and $(\Delta \alpha)_{\rm down}-(\Delta \alpha)_{\rm up}$, where the data for the up-sweep part of the pulse is subtracted to the data for the down-sweep. Raw data are shown in Fig. \ref{Stack_freq}.}
 	\label{Dif_Freq}
\end{figure}

The frequency dependence of the hysteresis is visible in Figs.~\ref{Stack_freq} and \ref{Dif_Freq}, where five different frequencies are compared. One clearly observes that the hysteresis associated with the superconductivity in field-polarized phase SC3 increases in size by lowering the ultrasonic frequency. The irreversibility field is highlighted by the dashed line $H_{SC}$. At the highest frequency of 388~MHz nearly no hysteretic behaviour is visible in $\Delta \alpha$ and in $\Delta v/v$ the hysteresis is not visible any more for $f\geq296$~MHz.\\

\clearpage


\section{Landau-Khalatnikov order parameter relaxation}

\begin{figure}[hb]
 	\includegraphics[width=1\linewidth]{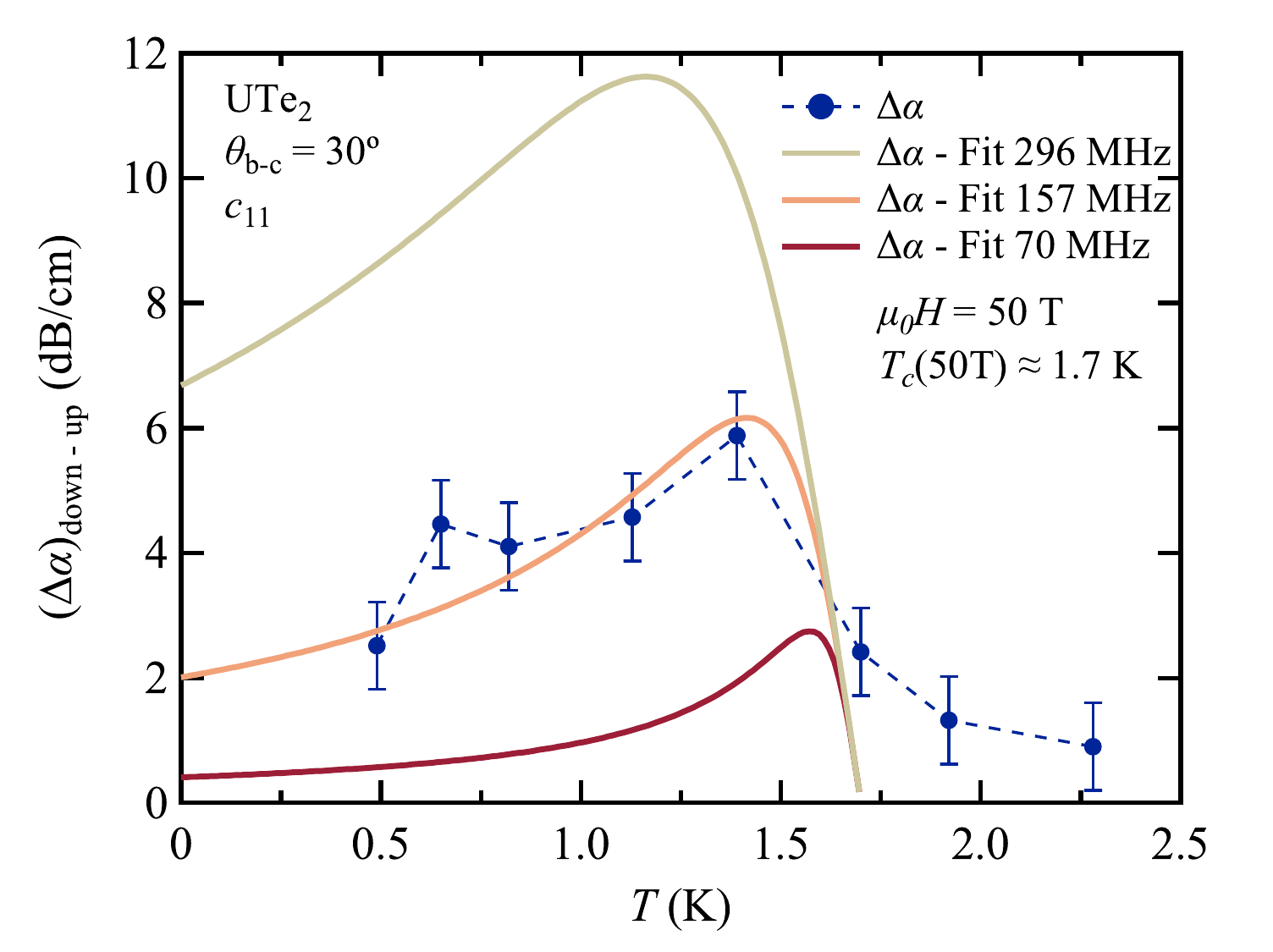}
 	\caption{Fit (orange line) of the attenuation hysteresis, $(\Delta \alpha)_{\rm down}-(\Delta \alpha)_{\rm up}$ (blue solide circles, see Fig. \ref{Dif_Hdep} and \ref{DiffupdwT}), measured at 50~T (applied with $\theta_{bc}=30^\circ$) and $f=157$~MHz in the $c_{11}$ mode, using Eq.~\ref{LKeq}. Fit parameters are $\tau_0=1.7\times 10^{-10}$~s and $T_{\rm c}=1.7$~K. Eq.~\ref{LKeq} is evaluated at 296~MHz (grey line) and 70~MHz (red line), using the same fit parameters.}
	\label{LKfit}
\end{figure}

Below superconducting $T_{\rm c}$ an order parameter appears. This order parameter can be driven out of equilibrium with a strain, and it will then relax to its equilibrium value after a characteristic relaxation time $\tau(T)$. This relaxation mechanism can cause attenuation inside the ordered phase, and it is usually called Landau-Khalatnikov (LK) mechanism. Following Ref.~\cite{Sigrist_2002,Ghosh_2022} we get the attenuation:

\begin{equation}
    \alpha(T,\omega) \propto \frac{\omega^2\tau(T)}{1+\omega^2\tau(T)^2}
    \label{LKeq}
\end{equation} with $\omega=2\pi f$, and $f$ the ultrasound frequency. Assuming $\tau(T) =\tau_0|T/T_{\rm c}-1|^{-1}$, we can fit the ultrasound data within this model, as shown in Fig.~\ref{LKfit}. The frequency dependence of the LK mechanism is also illustrated in Fig.~\ref{LKfit}. It is not in agreement with the measurements shown in Fig.~\ref{Stack_freq} and \ref{Dif_Freq}, as with increasing frequency the anomaly due to LK damping should increase, which is displayed in Fig.~\ref{LKfit}. This is not the case for the measured data in Fig.~\ref{Stack_freq}.

\newpage


\section{TAFF model}

The Thermally Assisted Flux Flow (TAFF) model is used to describe the dynamics of flux lines in type-II superconductors with strong pinning. The model starts with a pinning energy $U(T,H)$ which depends on temperature and applied magnetic field. The pinning energy dictate the dynamics of the flux lines. One can define a diffusion coefficient for the flux lines $c_{ii}^v\Gamma$, which can be phenomenologically related to the DC resistivity $\rho_e$ : $c_{ii}^v\Gamma=\rho_e (c^2/4\pi)$, with $\rho_e=\rho_0\textrm{exp}(-U/T)$, $\rho_0$ the normal state resistivity, and $c_{ii}^v$ the vortex lattice stiffness. The pinning will result in a modification of the sound velocity and attenuation which was derived by Pankert \textit{et al.} \cite{Pankert_1990}:

\begin{equation}
    \frac{\Delta v}{v}=\frac{1}{2\rho v^2}\frac{c_{ii}^v\omega^2}{\omega^2+(c_{ii}^v\Gamma k^2)^2}
    \label{Taffv}
    \end{equation}
    \begin{equation}
          \Delta \alpha = \frac{\omega^2}{2\rho v^3}\frac{(c_{ii}^v)^2\Gamma k^2}{\omega^2+(c_{ii}^v\Gamma k^2)^2} 
          \label{Taffa}
    \end{equation}
where $\omega$ is the frequency of the elastic wave with wave vector $k$ and $\rho$ is the density of the solid. At low $T$, when the vortices are rigidly pinned to the lattice, the sound velocity is increased due to the additional stiffness of the vortex lattice. When $T$ is increased, vortices are gradually depinned and attenuation rises at the depinning transition. When all vortices are thermally excited out of their pinning center, no coupling to the lattice remain. This model has been applied to describe the elastic properties of the mixed state of several superconductors quite successfully \cite{Pankert_1990,Wolf_1997,Wolf_1996}.

\begin{figure}[bh]
 	\includegraphics[width=1\linewidth]{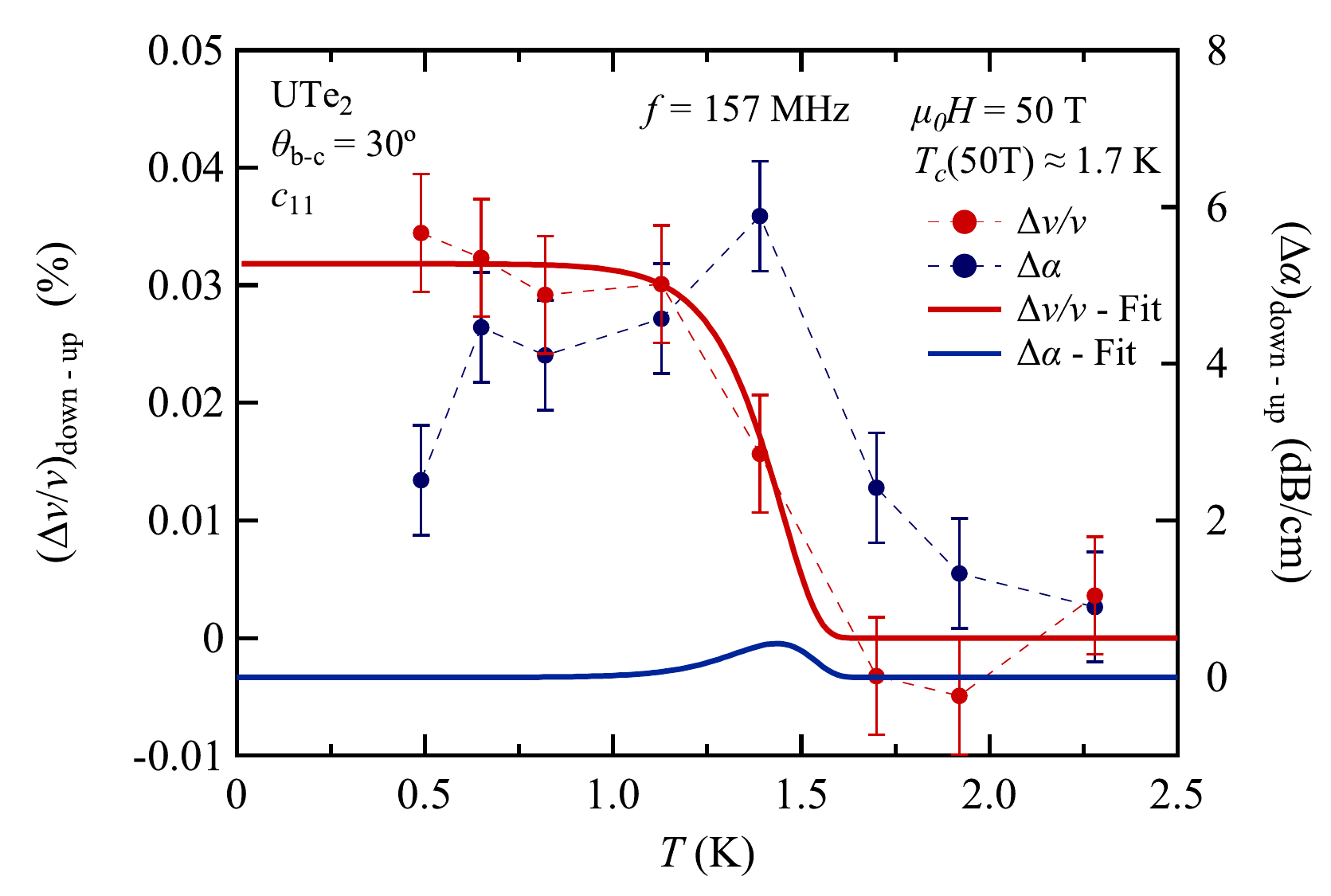}
 	\caption{Sound velocity (red) and attenuation (blue) in the $c_{11}$ mode as a function of $T$ at 50 T for $\theta_{b-c}=30^\circ$ (see Fig. \ref{DiffupdwT} for Data details). Red solid line is a fit using the TAFF model for the relative sound velocity described above. Fitting parameters are $U_0=192$~K, $\Delta U_0=41$~K and $n=0.5$. The vortex elastic constant is strongly reduced to $c_{ii}^v=0.07$~GPa, compared to the theoretically attended value at such a high magnetic field, which would result in a 30 time larger elastic constant. The fit is performed on the sound velocity data alone, such that the blue line (attenuation) is just the result of Eq.~\ref{Taffa}, using fit parameters determined on the sound velocity data. We use $\rho_0=60 ~\mu \Omega$.cm.}
	\label{TAFF}
\end{figure}

Here we try to apply the model to the elastic anomalies found in the field re-entrant superconducting phase of UTe$_2$. We assume $U(T,H)=U_0(1-T/T_{\rm c})^{n}$. In order to try to account for the width of the attenuation anomalies below $T_{\rm c}$, we use a Gaussian distribution of pinning energies $\Delta U_0$. We fit the sound velocity and attenuation hysteresis, respectively $(\Delta v/v)_{\rm down}-(\Delta v/v)_{\rm up}$ and $(\Delta \alpha)_{\rm down}-(\Delta \alpha)_{\rm up}$ from Fig. \ref{Dif_Hdep}, cut at fixed field in order to extract the temperature dependence of the ultrasound anomalies inside the SC-FP phase (see Fig.~\ref{DiffupdwT}). In Fig.~\ref{TAFF}, we report the ultrasound data at 50 T along with a fit using the TAFF equations~\ref{Taffv} and \ref{Taffa}. The sound velocity yields $U_0=192$~K, $\Delta U_0=41$~K and $n=0.5$ for an adapted vortex elastic constant $c_{ii}^v=0.07$~GPa (solid red line in Fig.~\ref{TAFF}). As discussed in the main text the vortex elastic constant $c_{ii}^v$ needs to be strongly reduced in order to model the ultrasound data, as for a magnetic field of $50$~T one attends roughly $2$~GPa which is roughly 30 times larger then the experimental determined change. The temperature dependence of the sound velocity is well reproduced, but the same cannot be said about the sound attenuation. Despite a sizable distribution $\Delta U_0/U_0 \approx 0.21$, the width of the attenuation anomaly and especially the amplitude of $\Delta \alpha$ cannot be reproduced.  Moreover, Eq.~\ref{Taffv} yields $\Delta v/v(T\rightarrow 0)\sim 10^{-2}$ at 50 T, while we measure $\Delta v/v \approx 3 \times 10^{-4}$, as seen in Fig.~\ref{DiffupdwT}, which leads to the fact that the absolute value of $c_{ii}^v$ was modified in order to model the data. Hence, the TAFF model works qualitatively, but fails to produce the observed excess attenuation for $T\ll T_{\rm c}$, and overestimates the sound velocity with respect to the measured attenuation at low $T$. Finally, Eq.~\ref{Taffv} and \ref{Taffa} cannot reproduce the large frequency dependence observed in Fig.~\ref{Stack_freq}. In the TAFF the anomalies increase with increasing frequency compared to the decrease observed in Fig.~\ref{Stack_freq}.
\newpage

\section{Heating}
\begin{figure}[b]
 	\includegraphics[width=1\linewidth]{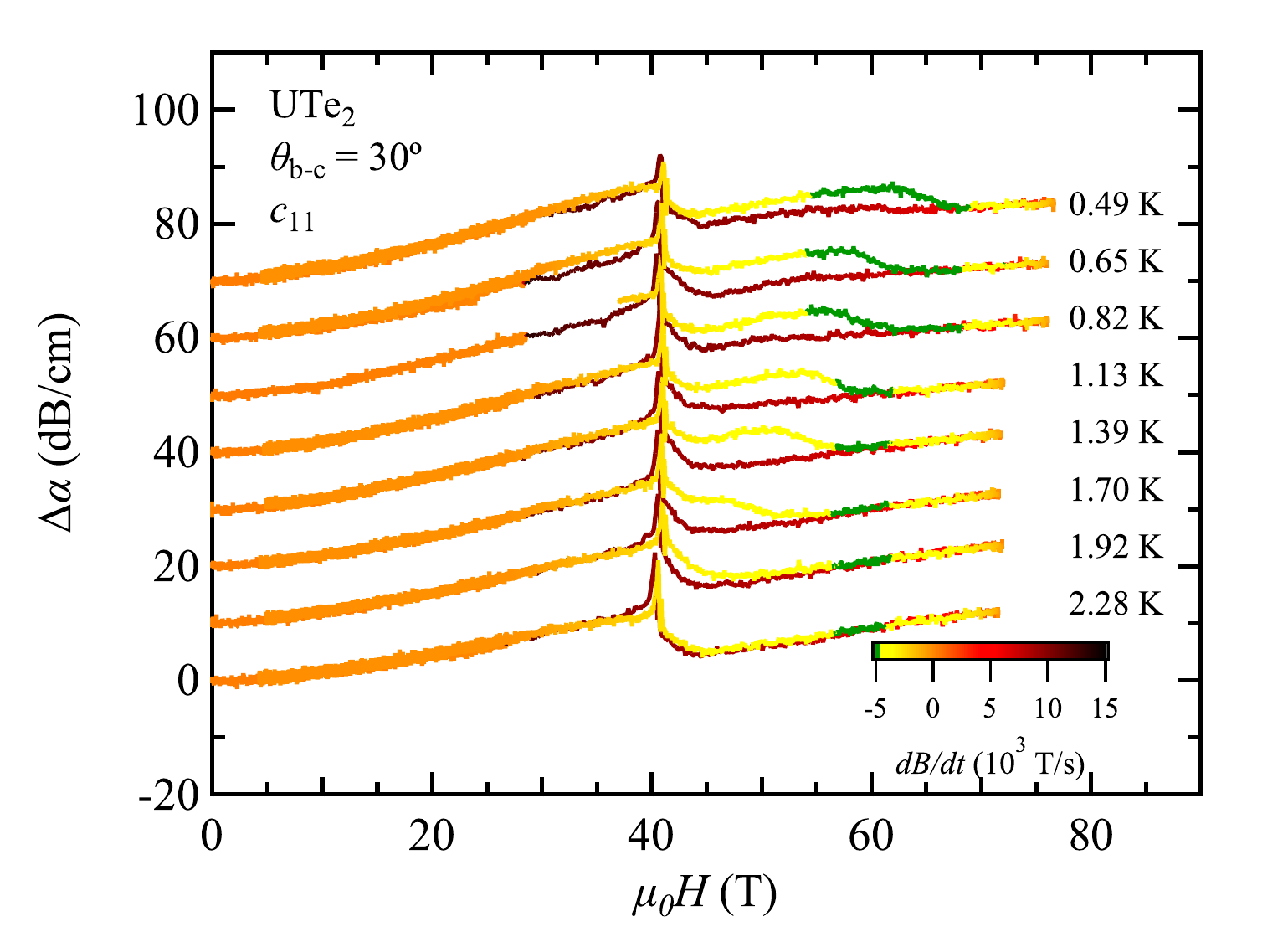}
	\caption{Variation in ultrasound attenuation $\Delta \alpha$ in the $c_{11}$ mode as a function of magnetic field, for different temperatures. The line color is coded with respect to $dB/dt$. Black indicates strong positive $dB/dt$, green indicates negative $dB/dt$ and orange indicates $dB/dt\sim 0$. $dB/dt$ is largest during the pulse of the inner coil from 28 to 80 T, with a maximum at the start of this short pulse (see Fig. \ref{specific_heat_field}a). The figure indicates that, while $dB/dt$ is significantly different between the up- and down-sweeps, and as such, may contribute to the size of the hysteresis, the opening of the hysteresis loop itself, at high fields for decreasing field, is not correlated to $dB/dt$, meaning it is not correlated to heating due to magnetic field induction.} 
	\label{heating}
\end{figure}
Here we discuss in more details the possible origins of heating effects during our pulsed field studies. Before we discuss this, it is important to say that the temperature before and after the pulsed magnetic field change is identical. For nearly every pulsed field measurement in this study the temperature is controlled via the vapor pressure of liquid \textsuperscript{4}He or \textsuperscript{3}He. Therefore, two possible heating effects can occur during the pulse. Either a Joule heating due to induced eddy currents by the rapid field change, or an intrinsic heat released by the sample due to some quasi adiabatic conditions when passing the first order metamagnetic transition. 

Regarding the Joule heating one needs to check the magnetic field change $\frac{dB}{dt}$. Fig.~\ref{heating} shows this as a color change on the attenuation measurements of $c_{\rm 11}$ originally displayed in the main text in Fig.~2. Dark red corresponds to a strong positive $\frac{dB}{dt}$ and yellow to a negative $\frac{dB}{dt}$. Usually for Joule heating the sign of $\frac{dB}{dt}$ is irrelevant, only its amplitude counts. The maximal change occurs at the onset of the second superconducting coil. This is within all measurements near 28~T, see Fig.~\ref{specific_heat_field}a for a detailed field profile. At this field between the up and down sweeps of the magnetic field no significant change is observed, which could explain the strong hysteresis inside the field polarized superconducting region. Additionally in Fig.~\ref{heating} in green the maximal negative change of $\frac{dB}{dt}<-5000$~T/s is displayed. In absolute terms its three times smaller than the maximal change, but it somehow coincidences at low temperature with the opening of the hysteresis associated to superconductivity. Nevertheless for higher temperatures this coincidence is not given and therefore can't explain an additional hysteresis opening due to some Joule heating.

\begin{figure*}[ht]
 	\includegraphics[width=0.47\linewidth]{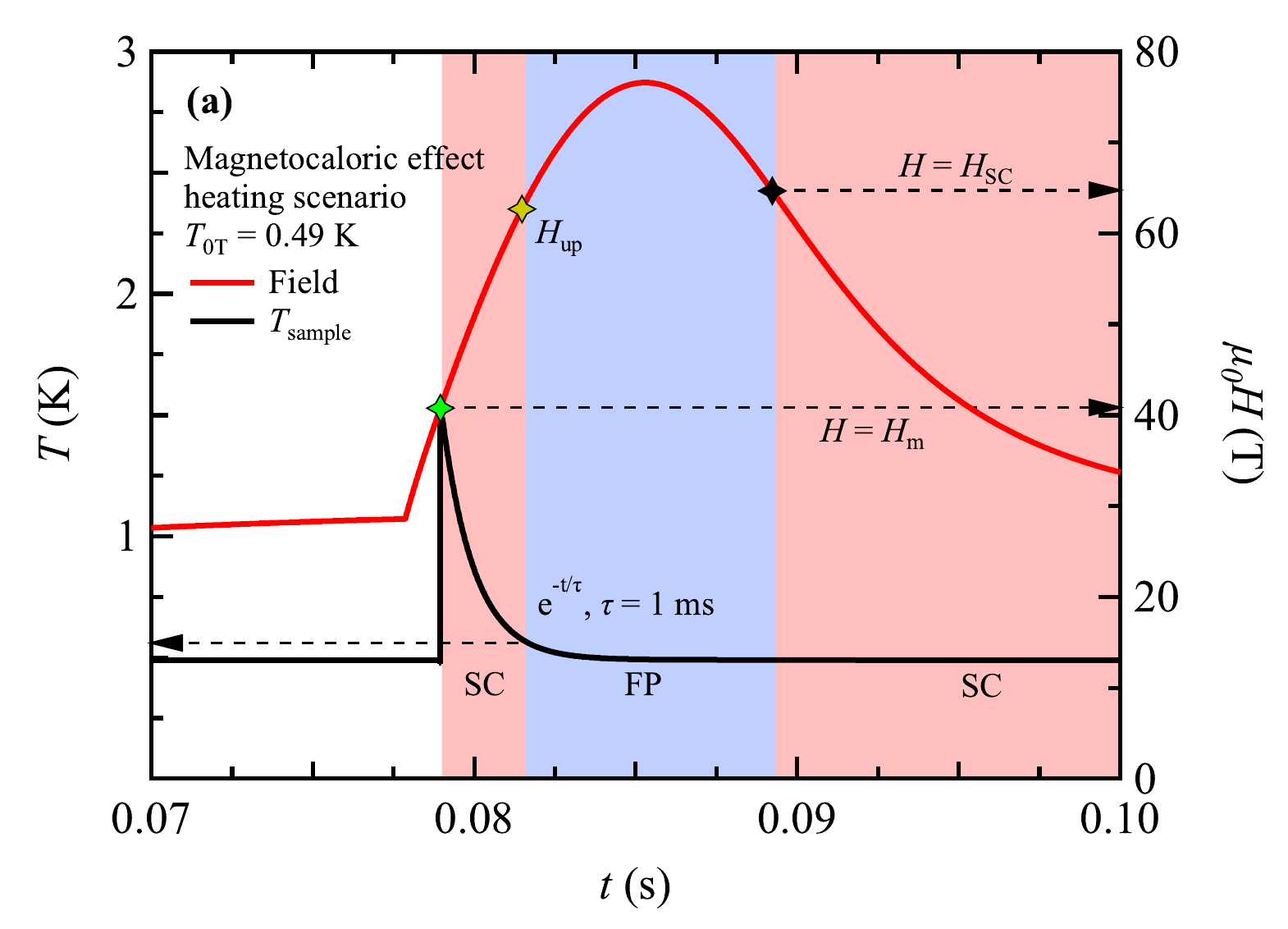}
 	\includegraphics[width=0.5\linewidth]{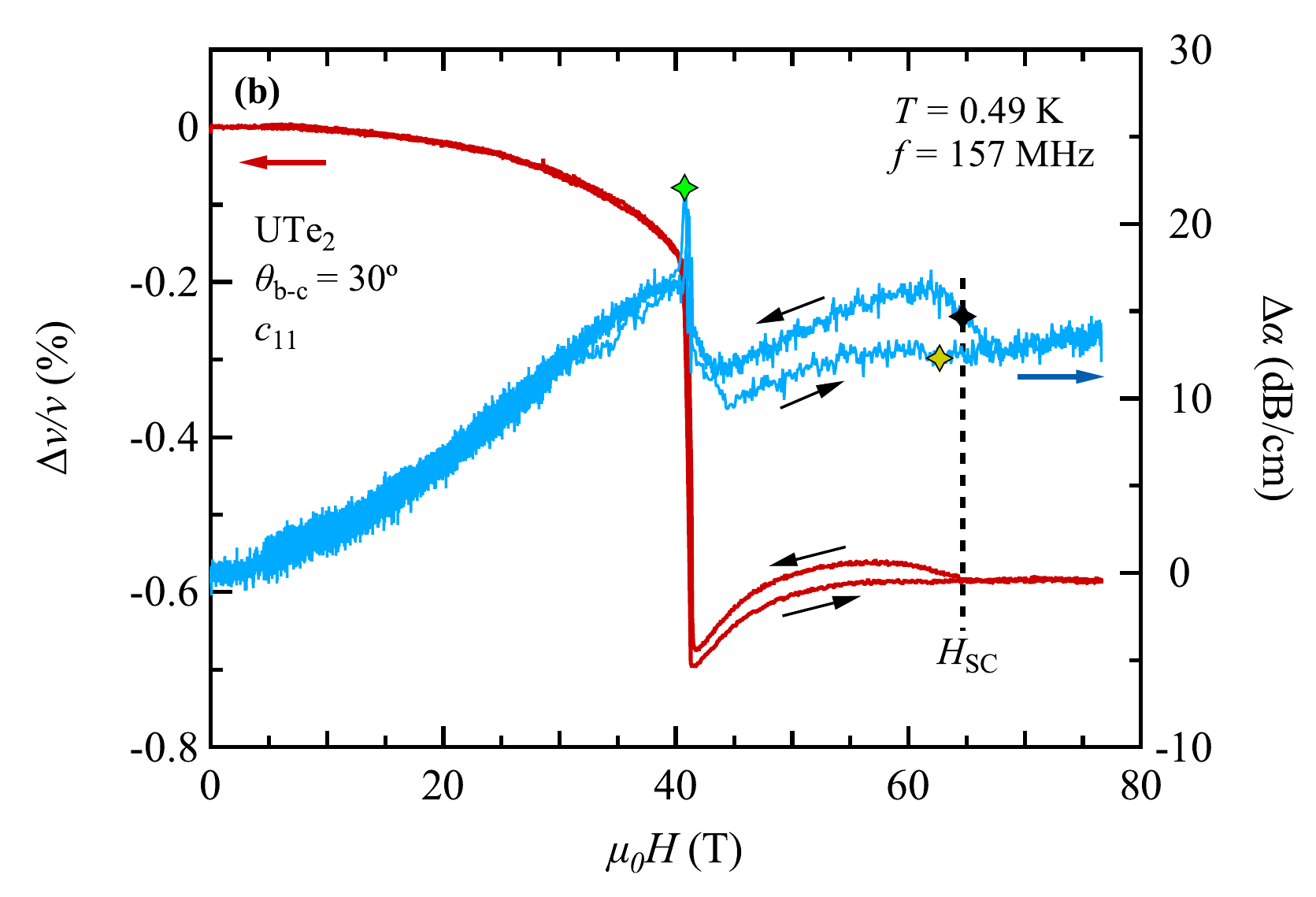}
 	\hfill
 	\includegraphics[width=0.47\linewidth]{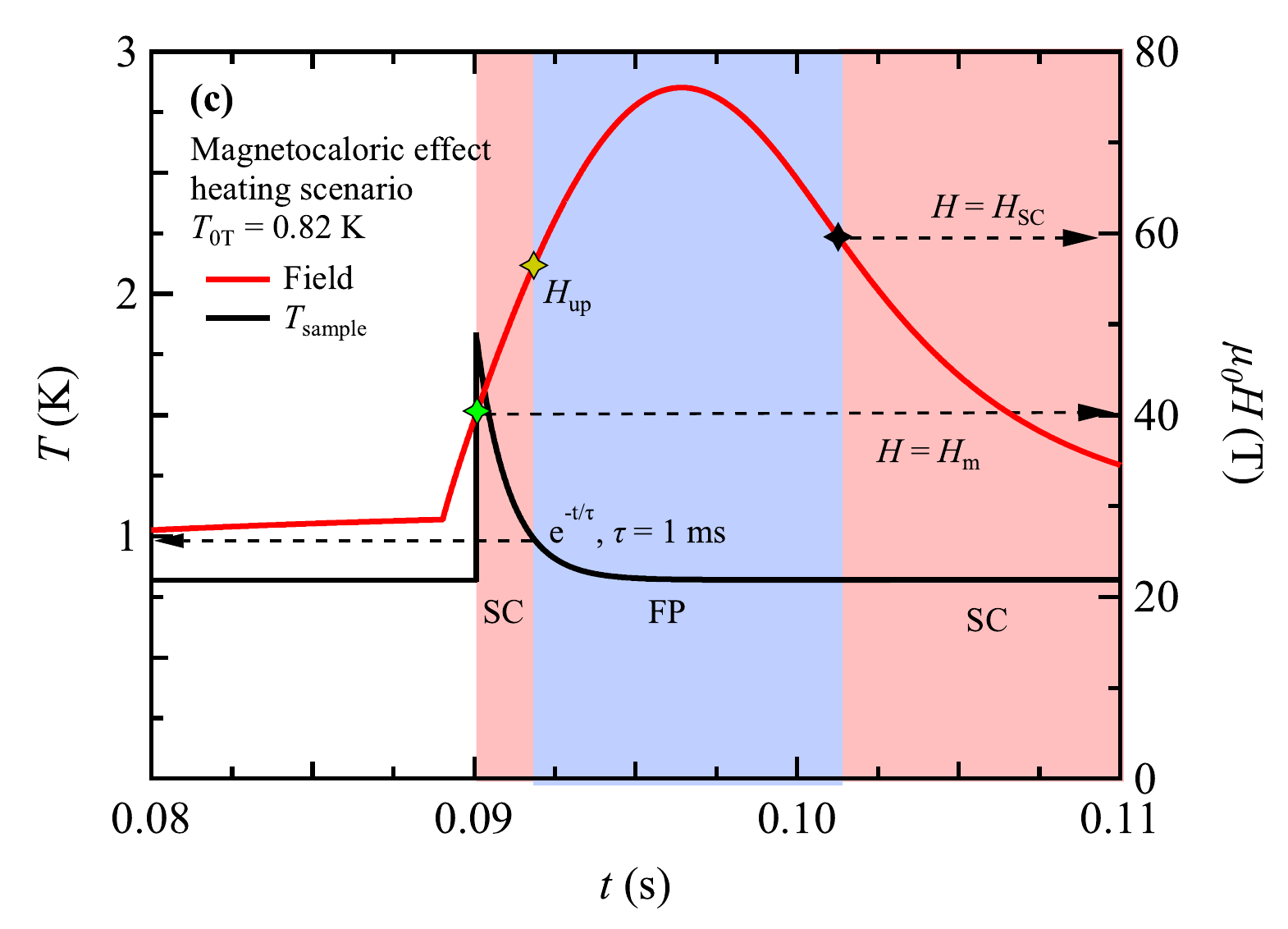}
 	\includegraphics[width=0.5\linewidth]{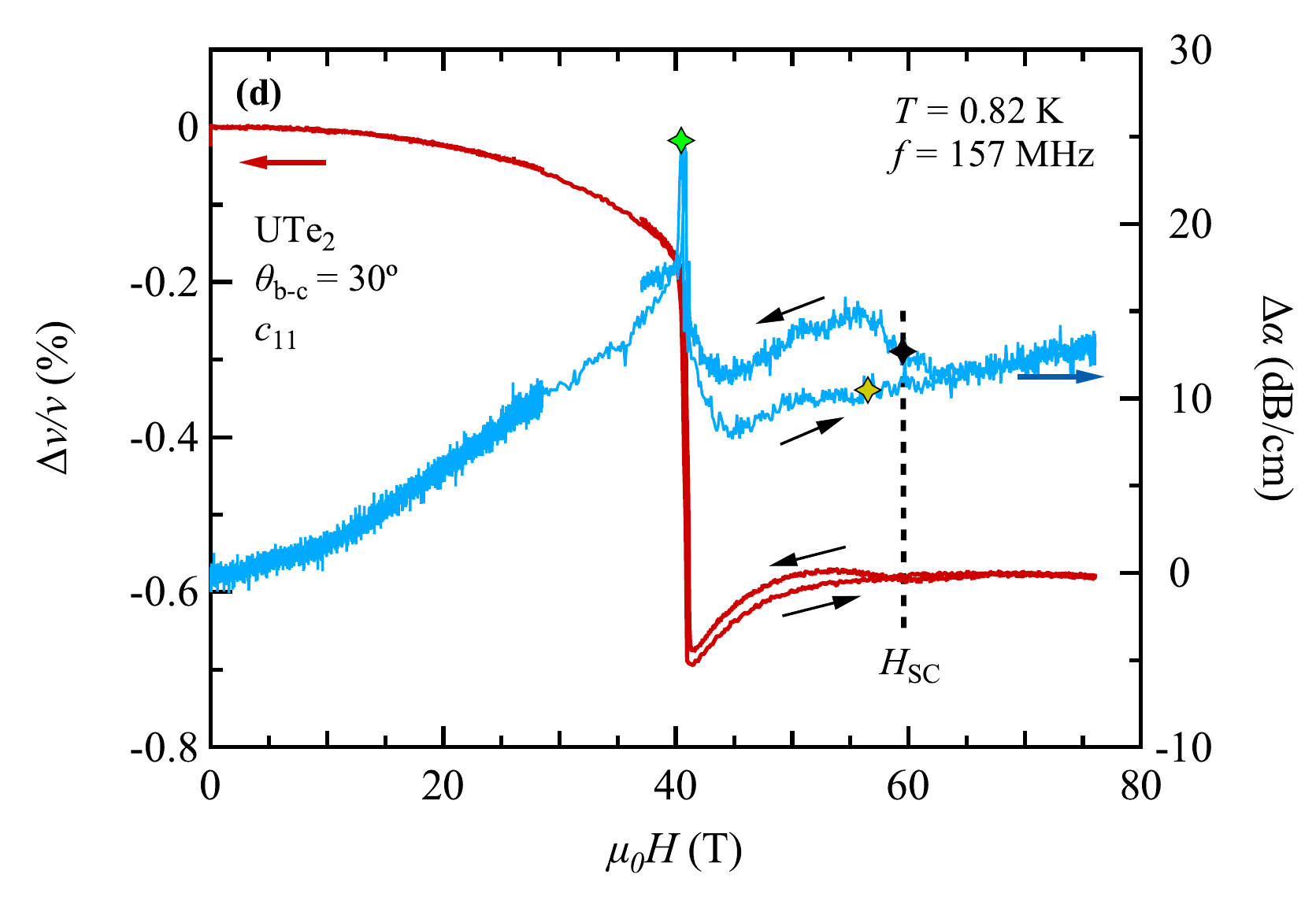}
 	\hfill
 	\includegraphics[width=0.47\linewidth]{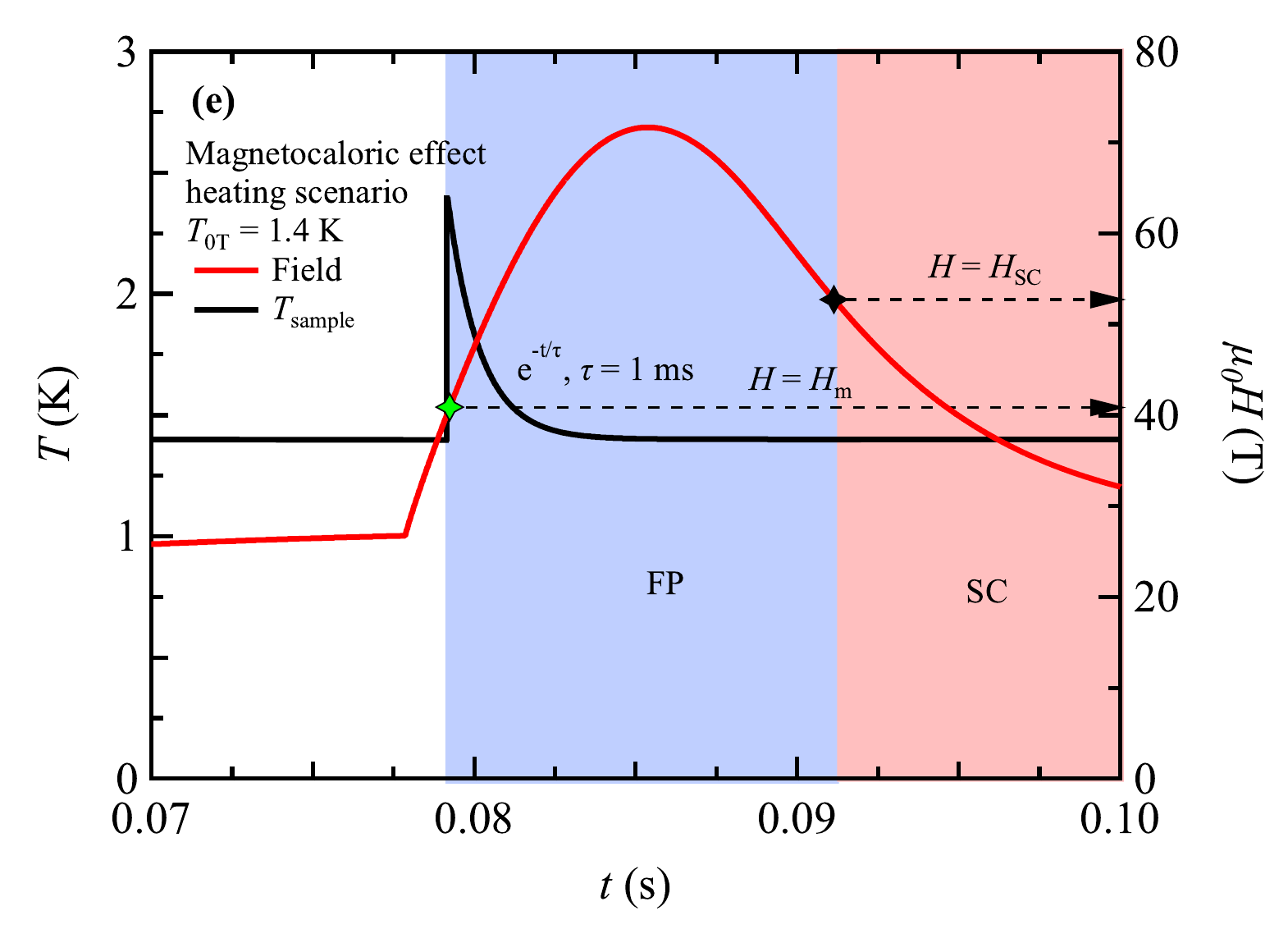}
 	\includegraphics[width=0.5\linewidth]{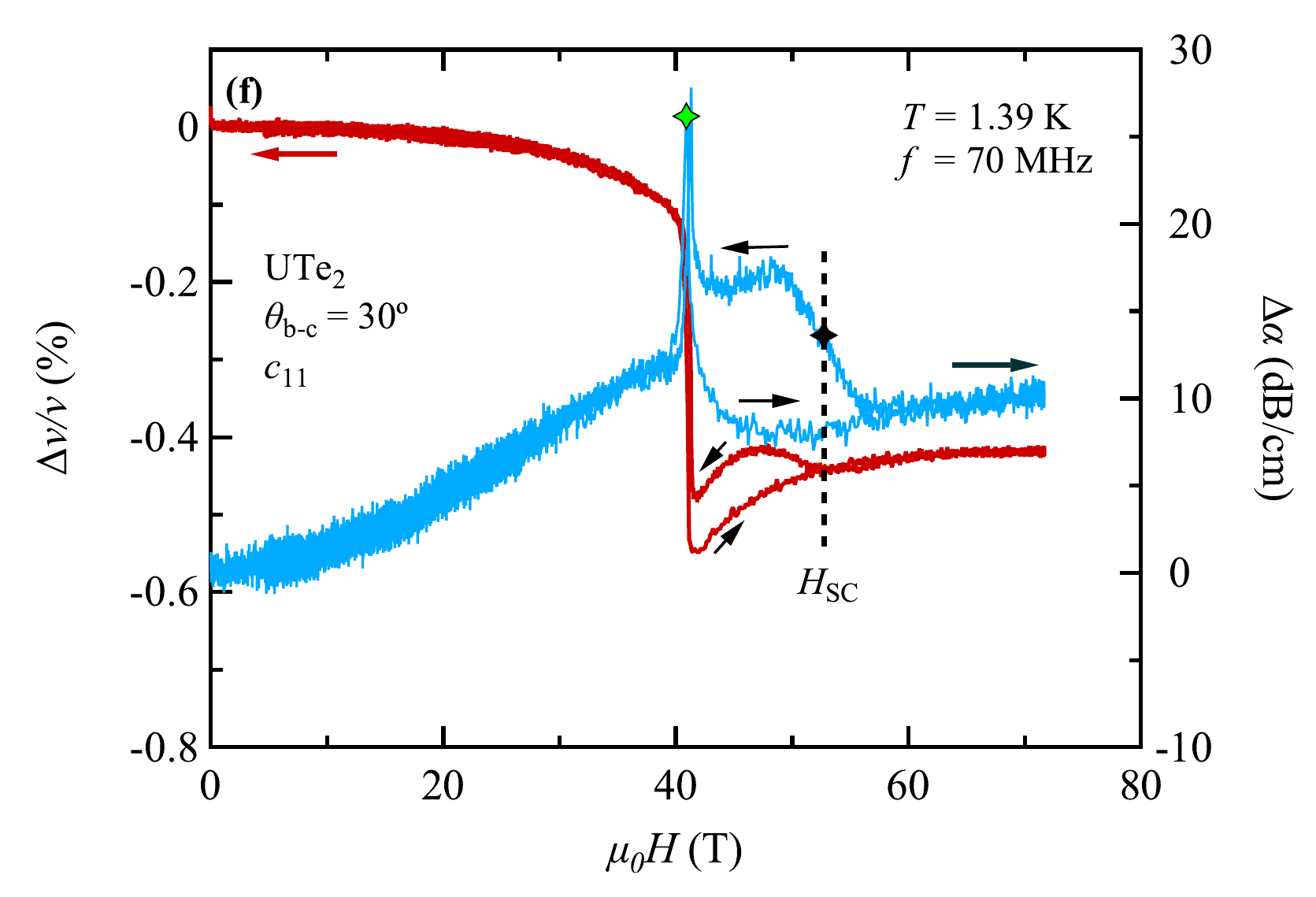}
 	\caption{The left hand panels (a,c,e) show a simulation of the sample temperature $T_{sample}$ in a quasi adiabatic scenario where the magnetocaloric effect at the metamagnetic transition increases the sample temperature by $\Delta T\approx1$~K. These simulations are performed for three zero field temperatures, corresponding to ultrasound measurements in $c_{11}$ which are shown on the right hand side panels (b,d,f). A thermal relaxation, with time constant $\tau$, to the bath is assumed by $e^{-t/\tau}$. In the same time dependent graphs the evolution of the respective magnetic field is displayed. The background color guides whether the superconducting state persists or if the field polarized state is entered.}
	\label{Simulation_temp}
\end{figure*}

\begin{figure}[ht]
 	\includegraphics[width=1\linewidth]{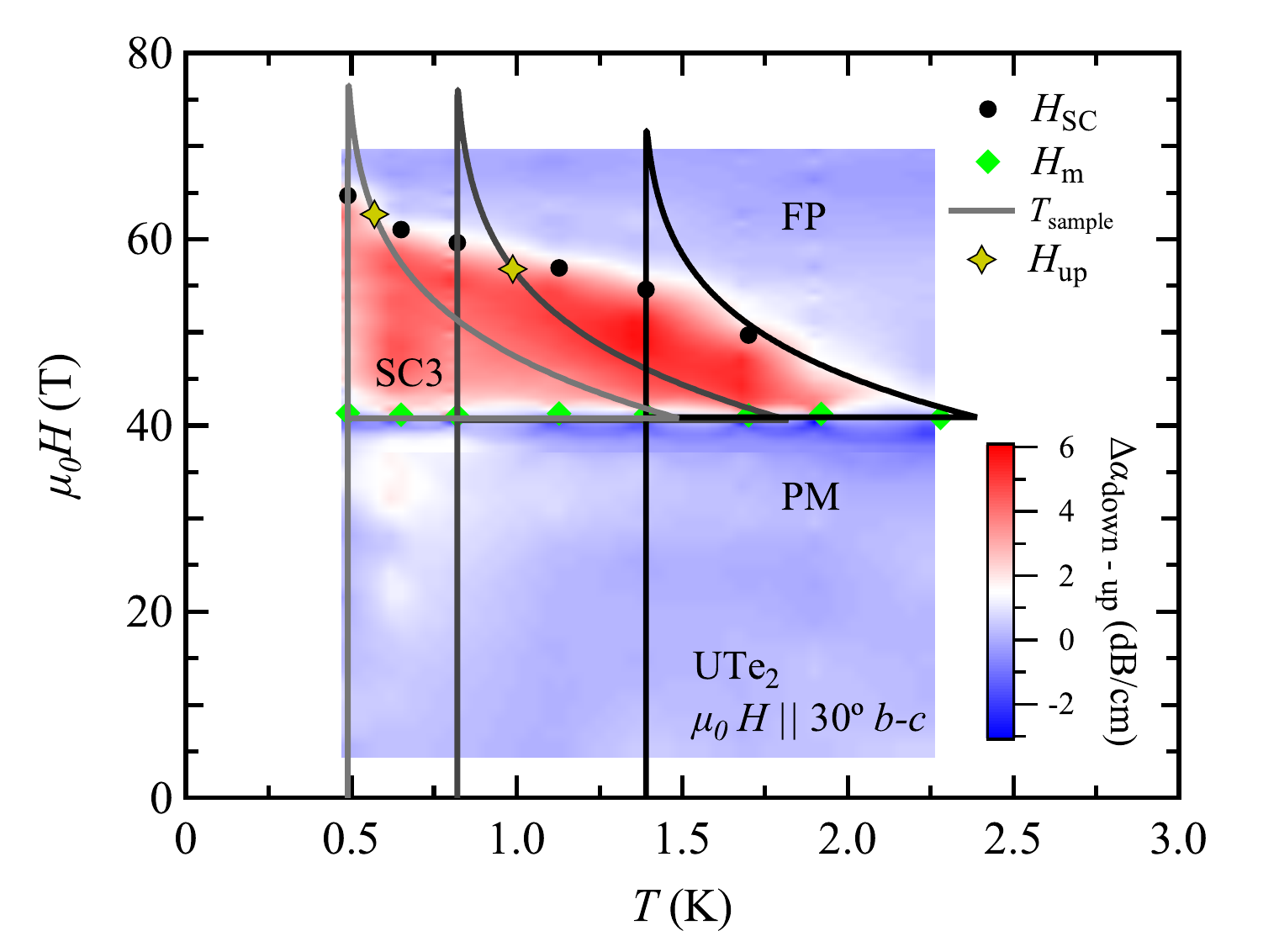}
 	\caption{The same phase diagram as in the main text is displayed. Additionally, in grey/black the field-temperature dependence of the sample temperature $T_{sample}$ for the magnetocaloric heating scenarios is added. The anomaly which is probably linked to superconductivity in the field increasing curve is highlighted by the yellow star $H_{\rm up}$. }
	\label{Phase_sc_simul}
\end{figure}

The influence of a possible magnetocaloric effect heating, which occurs when passing the metamagnetic transition is intrinsic to the physics of the first order metamagnetic transition in UTe$_2$ and can't be avoided. In our conditions, when measuring the superconducting state, we are within liquid helium \textsuperscript{3}He. No thermally insulating capsule is used in our set up, as one would use when measuring the magnetocaloric effect. This means that the thermal relaxation time $\tau$, until the sample reaches again thermal equilibrium with the bath must be shorter compared to previous measurements in specific thermally isolated, quasi adiabatic, conditions on the magnetocaloric effect at the metamagnetic transition in UTe$_2$, $\tau=10$~ms \cite{schonemann_sudden_2023} and $\tau>100$~ms \cite{imajo_thermodynamic_2019}. Due to a better thermal coupling to the bath for this set up one can assume a thermal relaxation time of $\tau \approx 1$~ms. A difference in thermal relaxation time when the sample is superconducting or in the normal state is neglected here. Both studies report comparable temperature increases when passing the metamagnetic transition, wherefore one can assume a strong temperature change of about $\Delta T =1$~K when passing the metamagnetic transition for all the temperatures measured here. In this minimal model these parameters of $\Delta T$ and $\tau$ lead to scenarios of the respective sample temperatures $T_{sample}$ evolution in time, which are displayed in Fig.~\ref{Simulation_temp}~(a,c,e) for three different zero field temperatures, 0.49~K, 0.82~K and 1.4~K. Additionally, the same graph shows the field evolution with time. Respective ultrasound measurements are shown on the right hand side in Fig.~\ref{Simulation_temp}~(b,d,f). For each bath temperature in both panels (left and right), the metamagnetic transition is highlighted by a green star. The superconducting anomaly at the irreversible field $H_{\rm SC}$ is shown in black. Note that within the simulation, when reentering the superconducting state in the down-sweep a stable sample temperature is reached again. New information are gained regarding the anomaly in the up-sweep measurements, yellow stars, as the corrected sample temperature at this instance can be extracted from the simulation. The sample temperature for these anomalies are indeed found to be slightly higher than the bath. This leads to the phase diagram in Fig.~\ref{Phase_sc_simul}, where the yellow stars are added compared to the anomalies which were already presented in the main text. This simple model seems to work quite well, as the temperature of the $H_{\rm up}$ anomaly locates rather well in the $H$-$T$ phase diagram. Additionally, as guide to the eye the simulated sample temperature evolution in the phase diagram is added. This can also explain the absence of any superconducting contribution in the up-sweep part of each measurement for temperatures above $T\geq 1.4$~K. 

Of course there are too many unknowns to determine the precise temperature increase and relaxation time by this method. In order to get the same location in the $H-T$ phase diagram for the $H_{\rm up}$ anomaly one could also lower the temperature jump at $H_{\rm m}$ and increase the relaxation time. In all cases the most important confirmation of such a model is that the temperature for the down-sweep data seems to be stable and back in equilibrium with the bath. Additionally it can't explain the different absolute values in attenuation, as all the down sweep measurements superimpose once the hysteresis opens, see Fig.~S11b.

\begin{figure}[ht]
 	\includegraphics[width=1\linewidth]{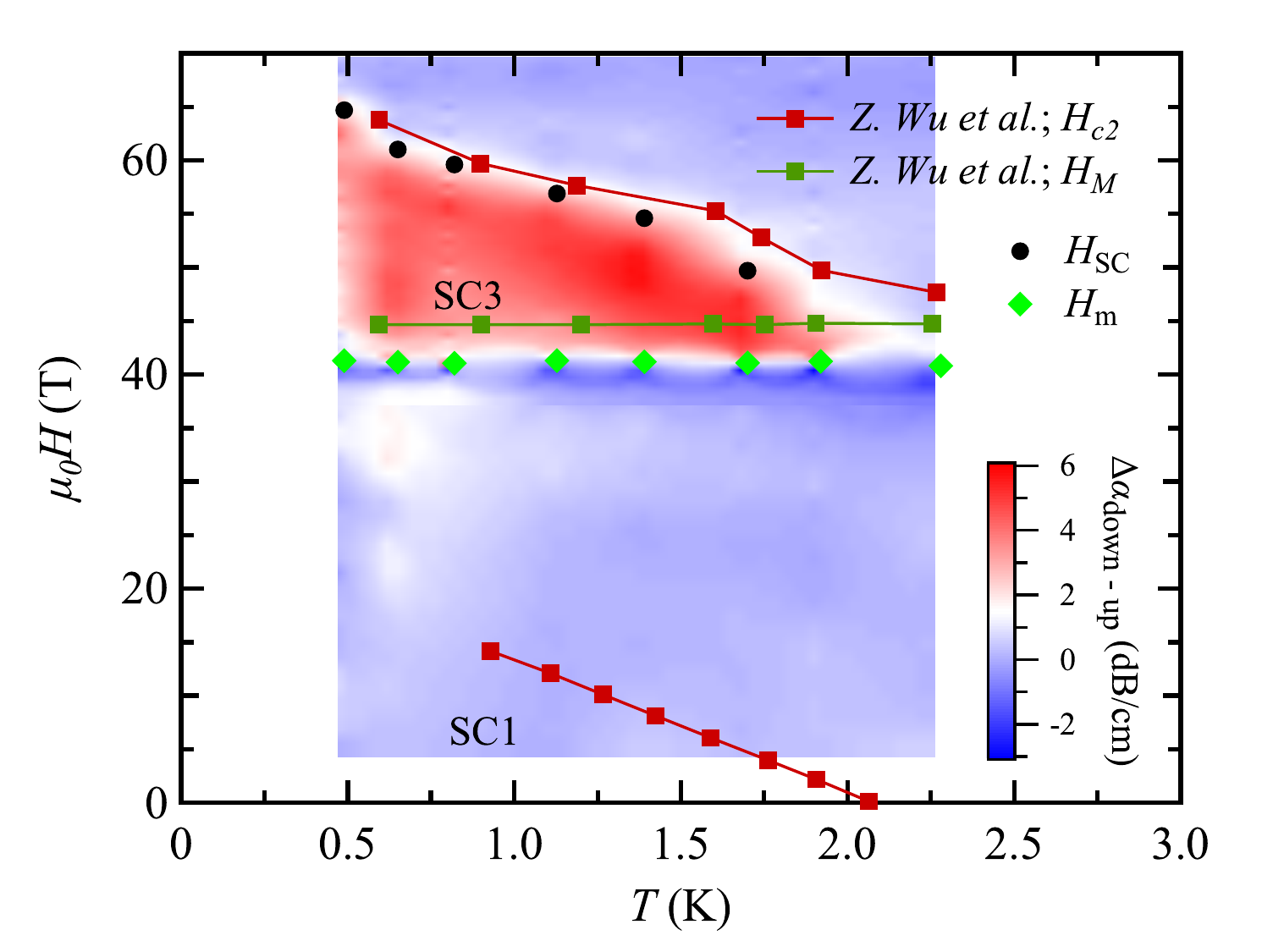}
 	\caption{Phase diagram of the field polarized superconducting phase (SC-FP) of UTe\textsubscript{2} for a magnetic field orientation of $30^\circ$ from $b$-$c$-axis in real space. From the relative sound velocity the jump in the field dependence at the metamagnetic transition (green) and the kink associated to the opening of the vortex flux line induced hysteresis (black) is displayed. The color plot highlights the amplitude of the hysteresis opening in attenuation and naturally highlights the expected SC-FP region. It is build upon data shown in Fig. \ref{Dif_Hdep}. Also plotted are phase boundaries from contactless surface conductivity PDO (Proximity Detector Oscillator) experiments in Ref.~\cite{wu_superconducting_2025} for a magnetic field orientated $35^\circ$ from $b$-$c$-axis. Red squares correspond to the mid-point of the PDO transition (local maximum in $df_{\rm PDO}/dB$), while green squares correspond to the metamagnetic transition \cite{wu_superconducting_2025}.}
	\label{Phase_sc_wu}
\end{figure}

\begin{figure}[ht]
 	\includegraphics[width=1\linewidth]{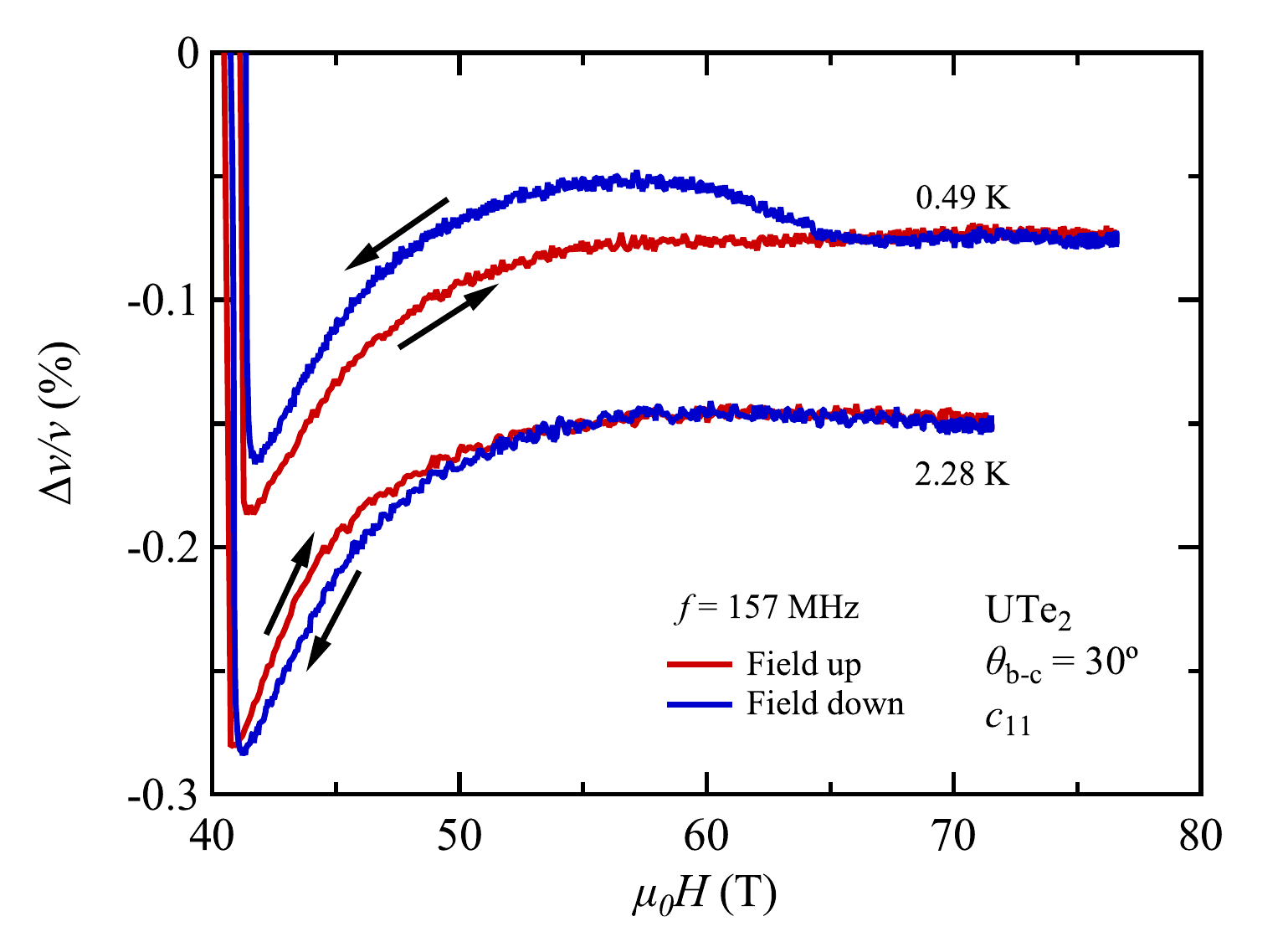}
 	\caption{A detailed view on the direction of the hysteresis in $\Delta v/v$ for two different temperatures is given. At 2.28~K above the superconducting critical temperature and at 0.49~K deep inside the superconducting state. The direction of the hysteresis changes, wherefore the red color stands for increasing and blue for decreasing magnetic field of the respective pulse.}
	\label{Hysteresis_direction}
\end{figure}

\begin{figure}[ht]
 	\includegraphics[width=1\linewidth]{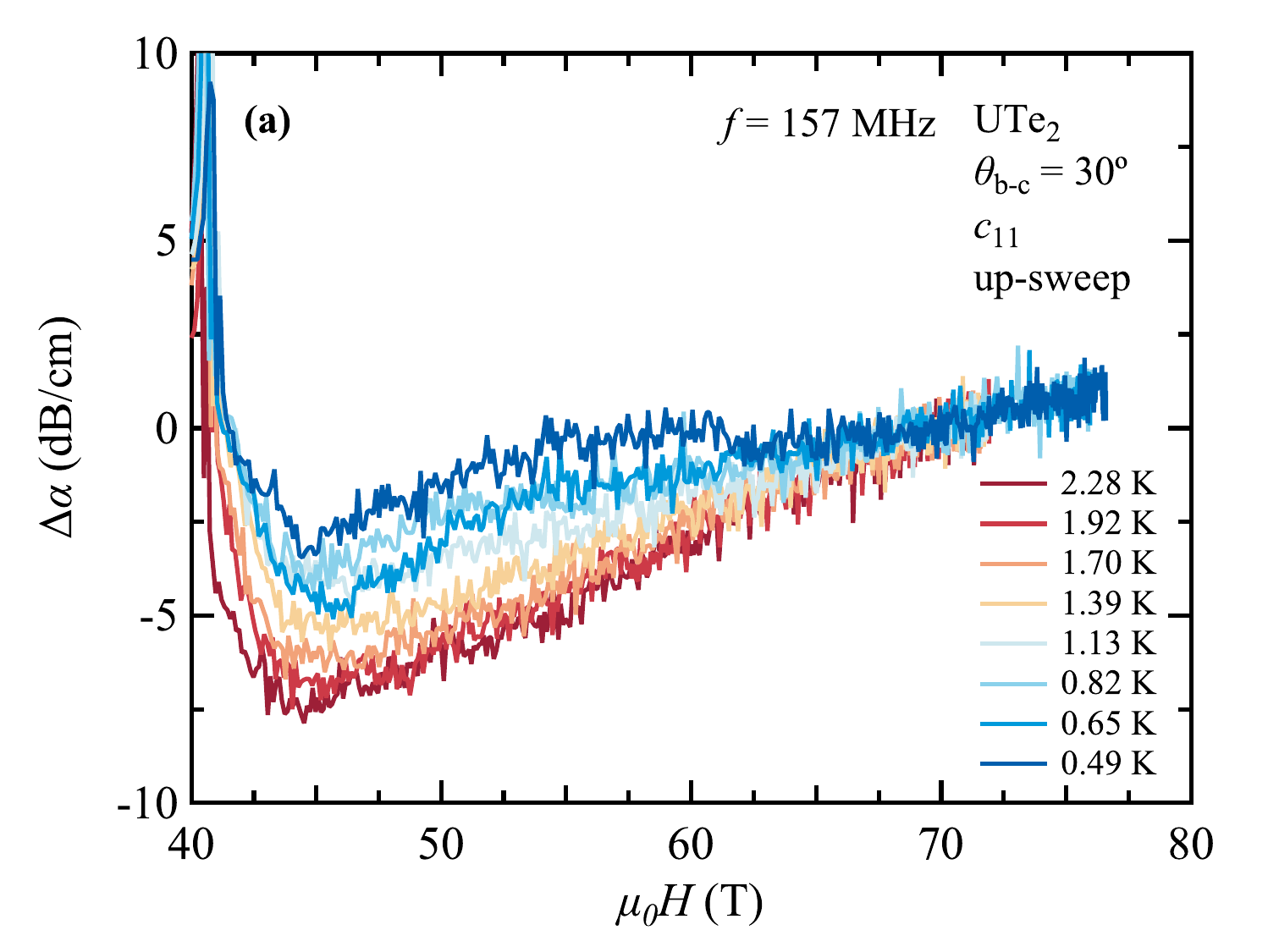}
\hfill
 	\includegraphics[width=1\linewidth]{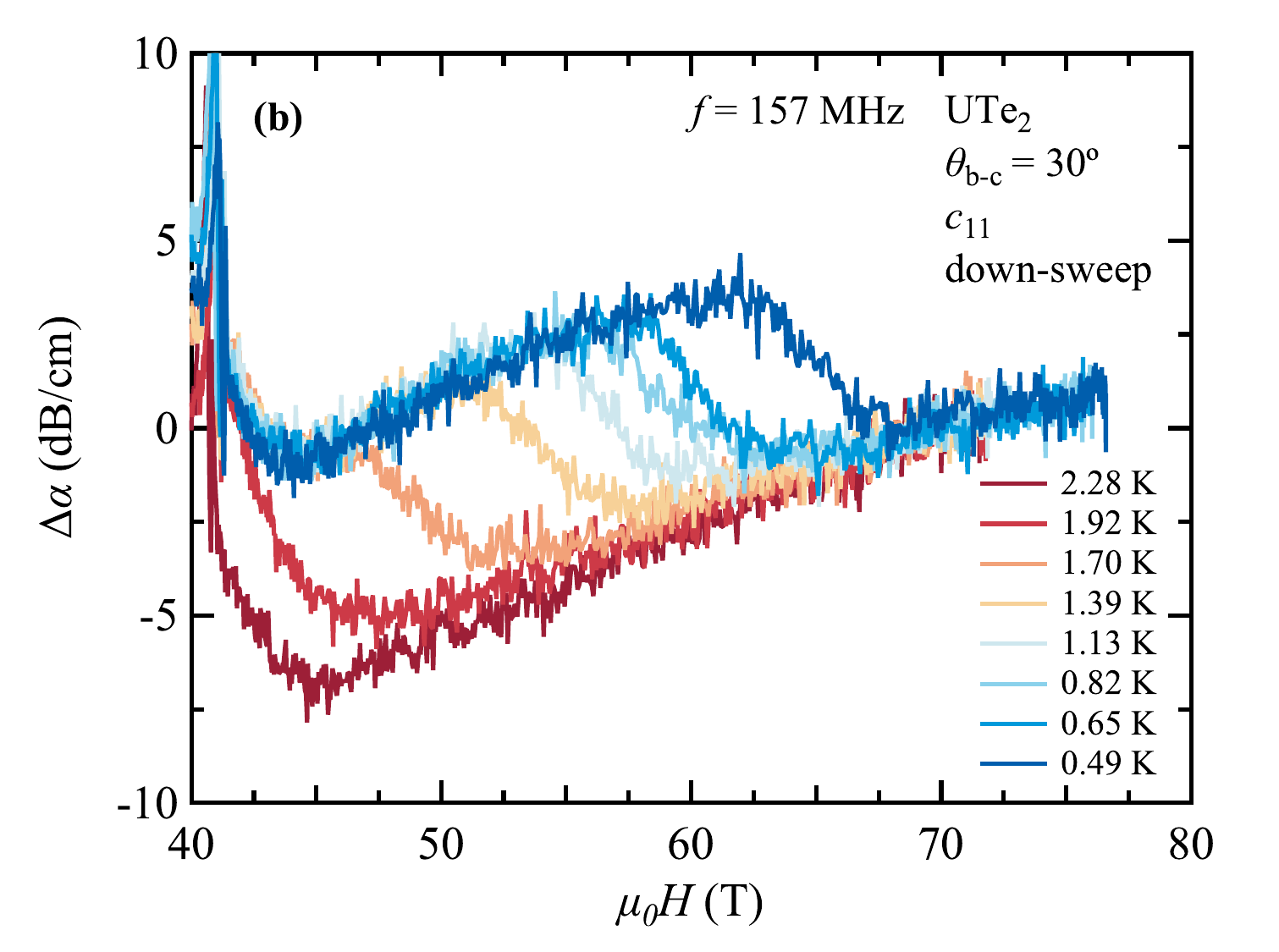}
 	\caption{Change in ultrasound attenuation $\Delta \alpha$ as a function of magnetic field measured in the $c_{11}$ mode for $\theta_{b-c}=30^\circ$, calculated with respect to the value measured at the highest field, for different temperatures. a) $\Delta \alpha$ for the up-sweep part of the magnetic pulse and b) for the decreasing part of the pulse. The anomalies are larger on the down-sweep part (panel b), but are also visible on the up-sweep (panel a).}
	\label{Deltaa_up_down}
\end{figure}

\begin{figure}[ht]
 	\includegraphics[width=1\linewidth]{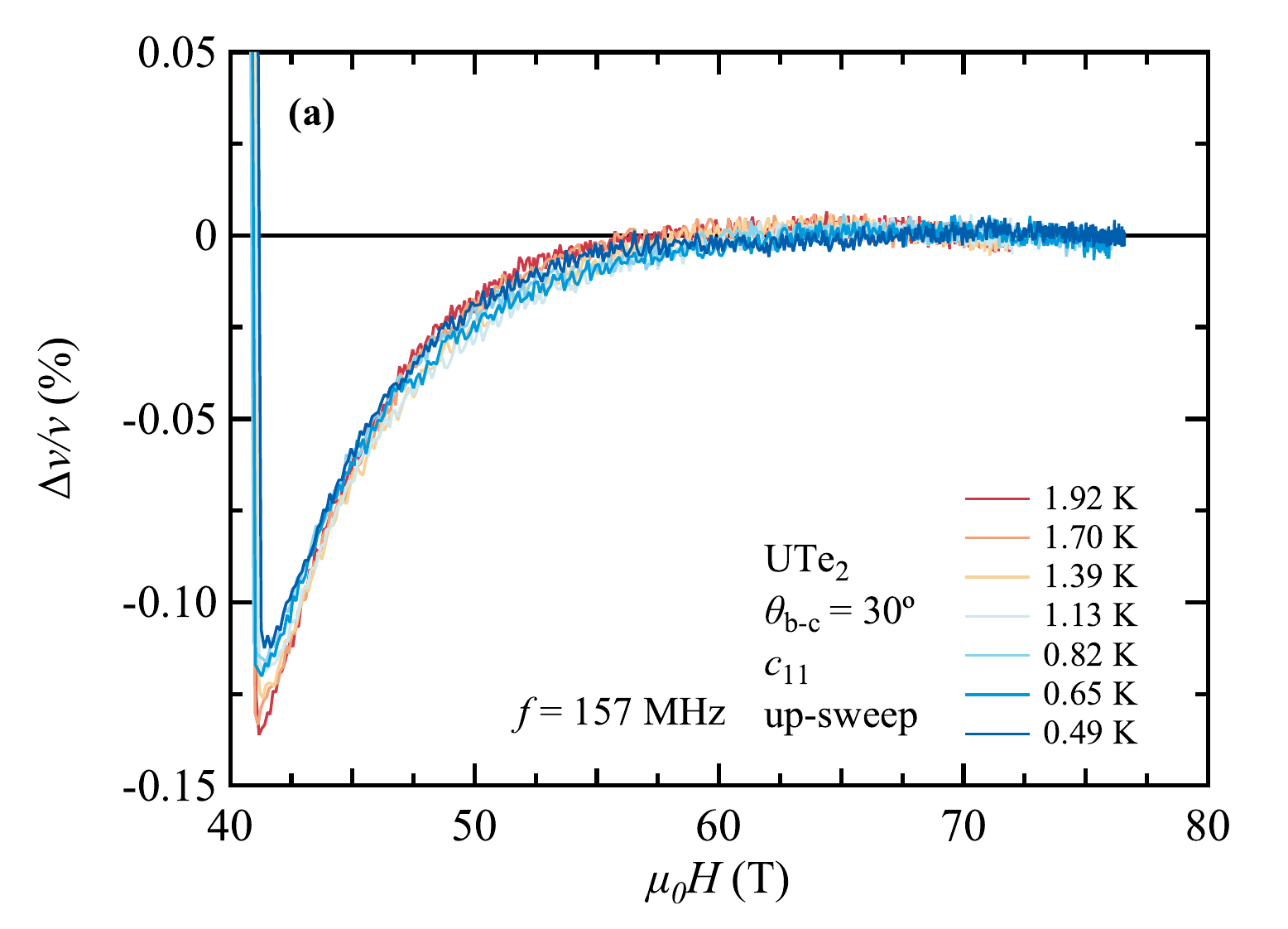}
\hfill
 	\includegraphics[width=1\linewidth]{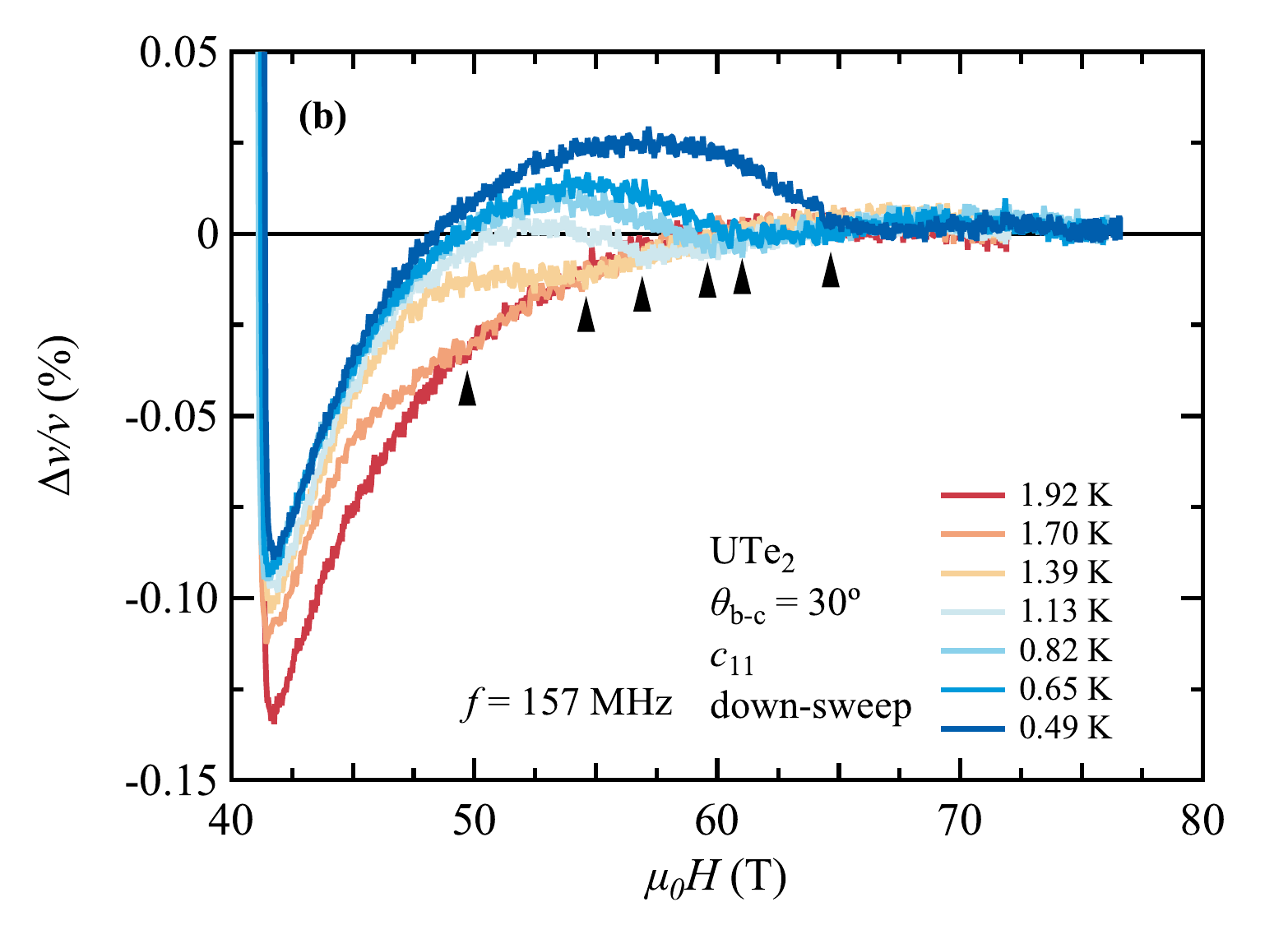}
 	\caption{Sound velocity change $\Delta v/v$ as a function of magnetic field measured in the $c_{11}$ mode for $\theta_{b-c}=30^\circ$, for different temperatures. Curves are shifted vertically so as to overlap at the highest measured fields. a) $\Delta v/v$ for the up-sweep part of the magnetic pulse and b) for the decreasing part of the pulse. The anomalies are larger on the down-sweep part (panel b), but are also visible on the up-sweep at the lowest measured $T$ (panel a).}
	\label{Deltav_up_down}
\end{figure}

 \begin{figure}[ht] 	
 \includegraphics[width=1\linewidth]{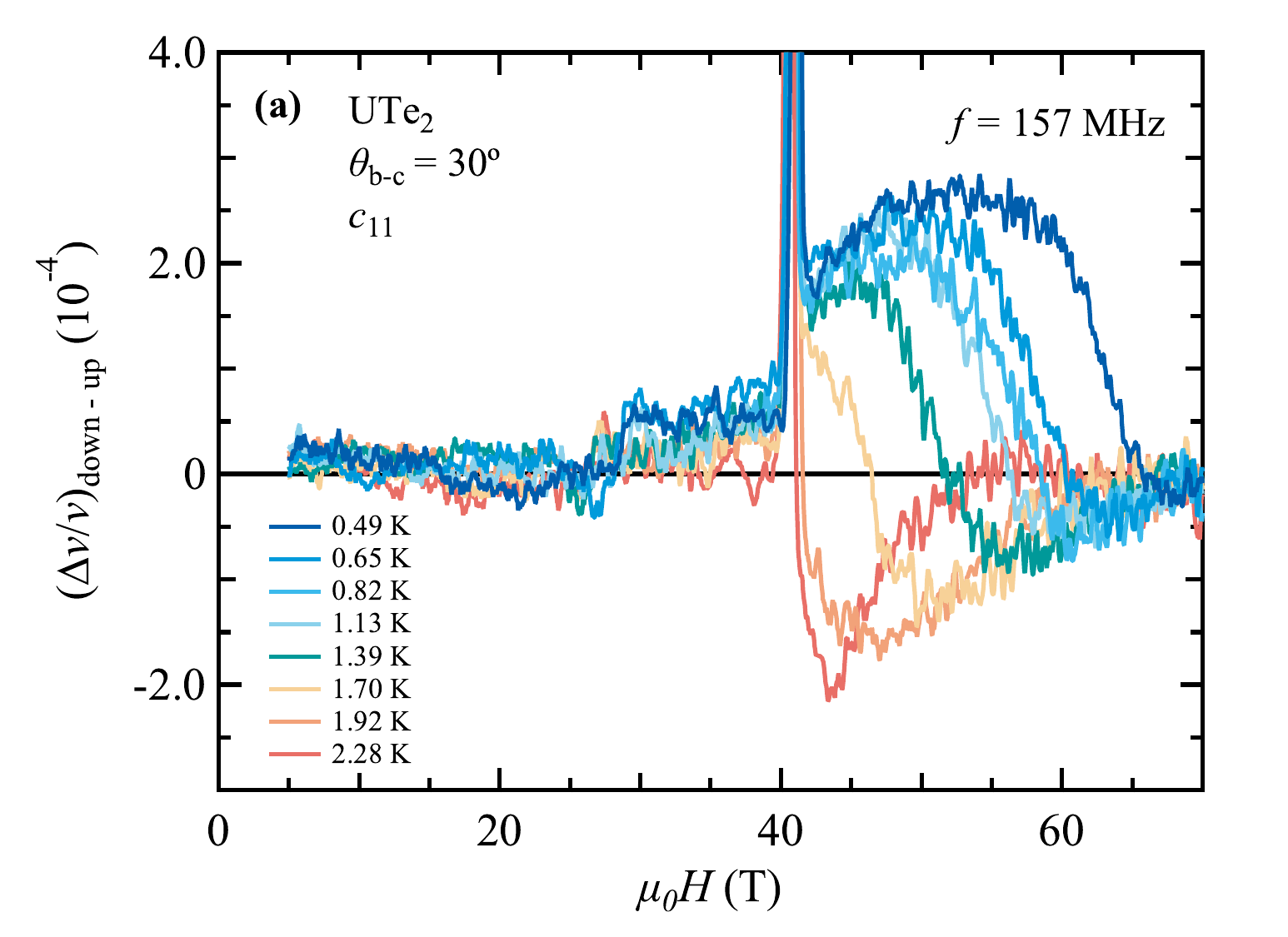}
\hfill
 	\includegraphics[width=1\linewidth]{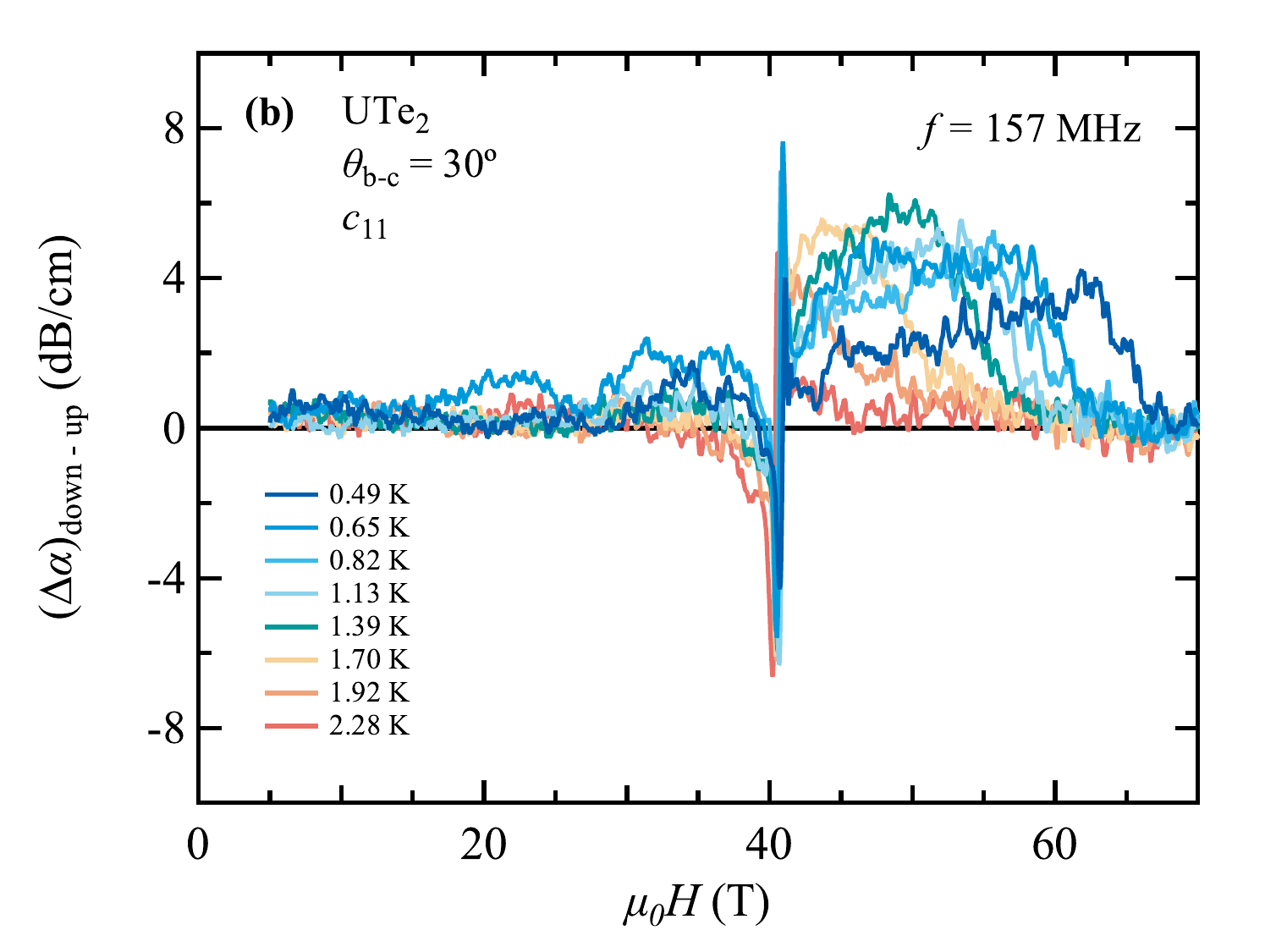}
 	\caption{Hysteresis loop in the field dependent ultrasound properties plotted as a) $(\Delta v/v)_{\rm down}-(\Delta v/v)_{\rm up}$ and b) $(\Delta \alpha)_{\rm down}-(\Delta \alpha)_{\rm up}$, for different temperatures. The data for the up-sweep part of the pulse is subtracted to the data for the down-sweep. The subtraction is made with the data shown in Fig. 1. Panel b is used to make the phase diagram of Fig. 3.}
	\label{Dif_Hdep}
\end{figure}

\begin{figure}[ht]
 	\includegraphics[width=1\linewidth]{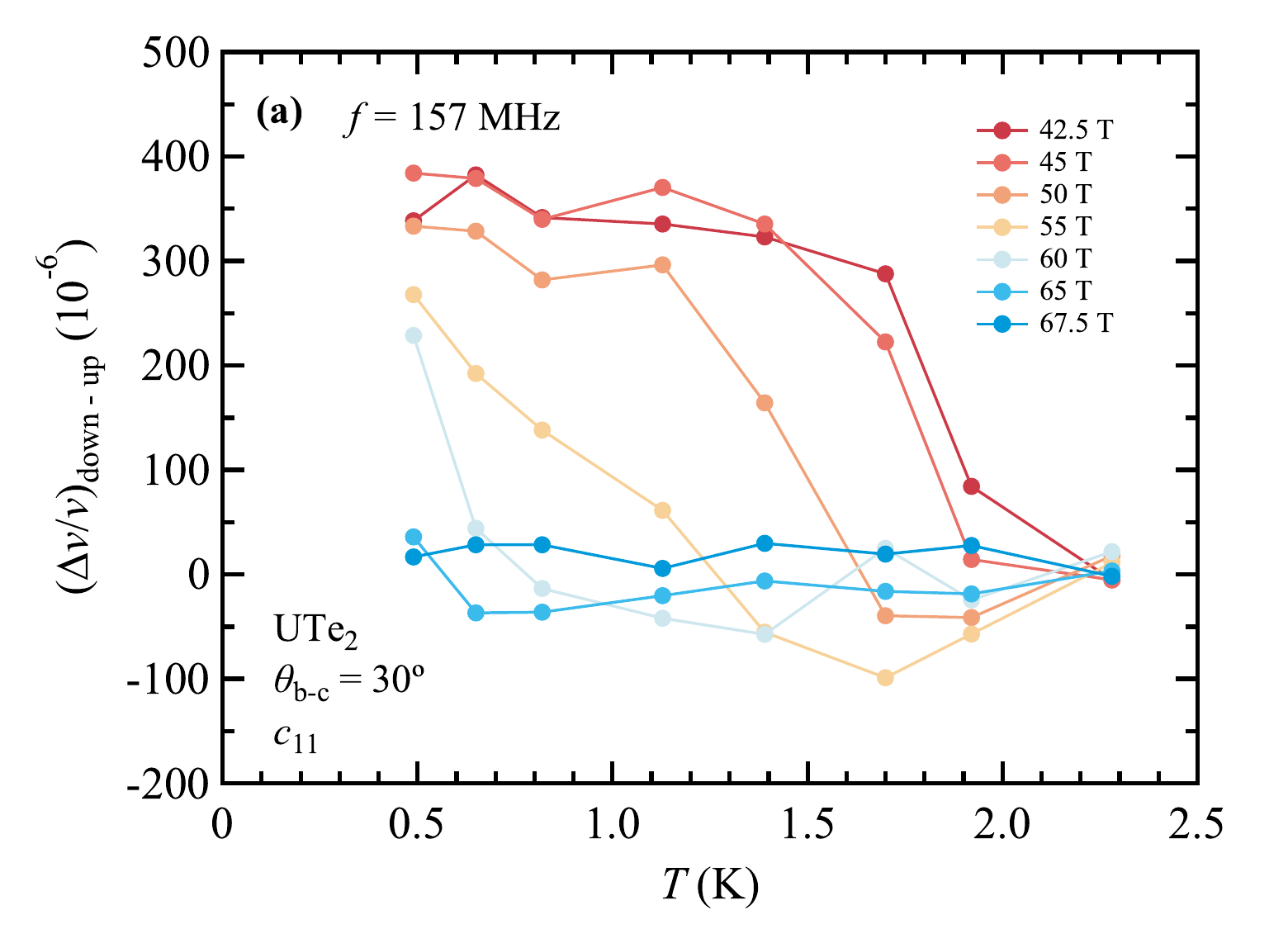}
\hfill
 	\includegraphics[width=1\linewidth]{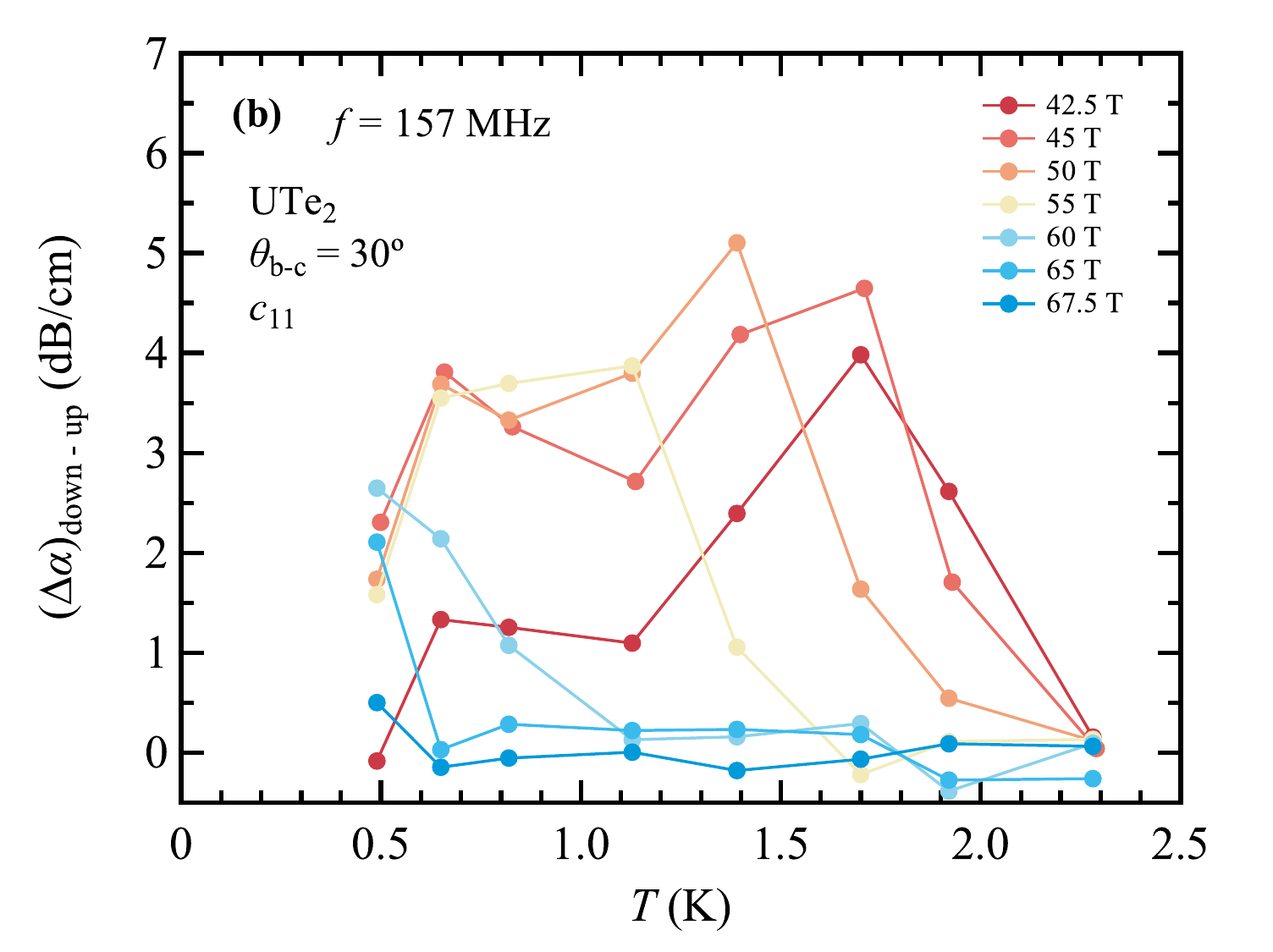}
 	\caption{Temperature dependence of the ultrasound anomalies, in the $c_{11}$ mode, inside the SC-FP phase, for different magnetic fields. Cuts at constant field are made to the data of Fig. \ref{Dif_Hdep}a (b) in order to produce panel a (b). As field is increased, the onset temperature of the ultrasound anomalies shifts towards lower $T$, in a similar way as superconducting $T_{\rm c}$. These data are used in Fig. \ref{LKfit} and \ref{TAFF} to perform fits with the LK and TAFF model respectively.}
	\label{DiffupdwT}
\end{figure}


\clearpage
\newpage


\end{document}